\documentclass[usenatbib]{mn2e}
\usepackage{natbib,graphicx,color}
\catcode`\@=11
\def\gsim{\ifmmode{\mathrel{\mathpalette\@versim>}}
    \else{$\mathrel{\mathpalette\@versim>}$}\fi}
\def\lsim{\ifmmode{\mathrel{\mathpalette\@versim<}}
    \else{$\mathrel{\mathpalette\@versim<}$}\fi}
\def\@versim#1#2{\lower 2.9truept \vbox{\baselineskip 0pt \lineskip
    0.5truept \ialign{$\m@th#1\hfil##\hfil$\crcr#2\crcr\sim\crcr}}}
\catcode`\@=12

\newcommand{\beq}{\begin{equation}}
\newcommand{\eeq}{\end{equation}}

\def\fzero{f_0}

\def\Wdm{W_{\rm dm}}
\def\Wzero{W_0}
\def\Wstar{W_{\rm *}}
\def\Wstarzero{W_{\rm *,0}}
\def\Wdmzero{W_{\rm dm,0}}

\def\phit{\phi_{\rm t}}

\def\rhalf{r_{\rm half}}

\newcommand{\rhostar}{\rho_*}
\newcommand{\rhostarzero}{\rho_{*,0}}
\newcommand{\rhostartilde}{\tilde{\rho}_*}

\newcommand{\Mstar}{M_*}

\newcommand{\Mdm}{M_{\rm dm}}
\newcommand{\Mdmvir}{M_{\rm dm,vir}}
\newcommand{\Mdmt}{M_{\rm dm,t}}

\newcommand{\ra}{r_{\rm a}}
\newcommand{\rarad}{r_{\rm a,rad}}
\newcommand{\ratan}{r_{\rm a,tan}}
\newcommand{\ratilde}{\tilde{r}_{\rm a}}
\newcommand{\rt}{r_{\rm t}}
\newcommand{\rttilde}{\tilde{r}_{\rm t}}
\newcommand{\rs}{r_{\rm s}}
\newcommand{\rvir}{r_{\rm vir}}

\newcommand{\rc}{r_{\rm c}}

\newcommand{\vr}{v_{\rm r}}
\newcommand{\vt}{v_{\rm t}}
\newcommand{\vx}{v_{x}}
\newcommand{\vy}{v_{y}}

\newcommand{\sigmaV}{\sigma_{\rm V}}
\newcommand{\sigmaK}{\sigma_{\rm K}}
\newcommand{\sigmar}{\sigma_{\rm r}}
\newcommand{\sigmat}{\sigma_{\rm t}}

\newcommand{\xihalf}{\xi_{\rm half}}

\newcommand{\Trad}{T_{\rm rad}}
\newcommand{\Ttan}{T_{\rm tan}}
\def\Ttheta{T_{\vartheta}}
\def\Tphi{T_{\varphi}}
\newcommand\sgth{\sigma_{\vartheta}}
\newcommand\sgthsq{\sigma_{\vartheta}^2}
\newcommand\sgph{\sigma_{\varphi}}
\newcommand\sgphsq{\sigma_{\varphi}^2}
\newcommand\sgr{\sigma_r}
\newcommand\sgrsq{\sigma_r^2}
\def\Sigmastar{\Sigma_*}

\newcommand{\rhodm}{\rho_{\rm dm}}
\newcommand{\rhodmzero}{\rho_{\rm dm,0}}

\def\d{{\rm d}}

\newcommand{\bey}{\begin{eqnarray}}
\newcommand{\eey}{\end{eqnarray}}
\newcommand{\Myr}{\, {\rm Myr} }
\newcommand{\Gyr}{\, {\rm Gyr} }

\newcommand{\pc}{\, {\rm pc} }
\newcommand{\kpc}{\, {\rm kpc} }
\newcommand{\Msun}{M_\odot}
\newcommand{\LV}{L_V}

\newcommand{\grad}{{\bf \nabla}}

\def\ltsima{$\; \buildrel < \over \sim \;$}
\def\simlt{\lower.5ex\hbox{\ltsima}}
\def\gtsima{$\; \buildrel > \over \sim \;$}
\def\simgt{\lower.5ex\hbox{\gtsima}}
\newcommand{\fmmm}[1]{\mbox{$#1$}}
\newcommand{\scnd}{\mbox{\fmmm{''}\hskip-0.3em .}}

   \title[Dark matter in NGC~2419]
         {Do globular clusters possess Dark Matter halos?\\ A case study in NGC~2419}
\author[]{}

   \author[Ibata et al.]
          {\parbox{\textwidth}{R. Ibata$^1$\thanks{E-mail:rodrigo.ibata@astro.unistra.fr}, 
           C. Nipoti$^2$, 
           A. Sollima$^3$,
           M. Bellazzini$^4$,
           S.C. Chapman$^5$ and
           E. Dalessandro$^2$}\vspace{0.4cm}\\ 
\parbox{\textwidth}{$^1$Observatoire Astronomique, Universit\'e de Strasbourg, CNRS, 11, rue de l'Universit\'e, F-67000 Strasbourg, France
\\$^2$Department of Physics and Astronomy, viale Berti Pichat, 6/2, I-40127 Bologna, Italy
           \\ $^3$INAF - Osservatorio Astronomico di Padova, vicolo dell'Osservatorio 5, 35122, Padova, Italy
           \\ $^4$INAF-Bologna Astronomical Observatory, 
                       via Ranzani 1, I-40127 Bologna, Italy
\\ $^5$Institute of Astronomy, Madingley Road, Cambridge CB3 0HA, UK
}}

\date{September 18, 2012}
\pubyear{2012}

\begin{document} 
\maketitle

\begin{abstract}
We use recently published measurements of the kinematics, surface
brightness and stellar mass-to-light ratio of the globular cluster
NGC~2419 to examine the possibility that this Galactic halo satellite
is embedded in a low-mass dark matter halo. NGC~2419 is a promising
target for such a study, since its extreme Galactocentric distance and
large mass would have greatly facilitated the retention of dark
matter. A Markov-Chain Monte Carlo approach is used to investigate
composite dynamical models containing a stellar and a dark matter component. We
find that it is unlikely that a significant amount of dark matter ($\simlt6\%$ of the luminous mass inside
the tidal limit of the cluster) can be present if the
stars follow an anisotropic Michie model and the dark matter a
double power law model.
However, we find that more general models, derived using
a new technique we have developed to compute 
non-parametric solutions to the spherical Jeans equation,
suggest the presence of a significant dark matter fraction (approximately twice the stellar mass).
Thus the presence of a dark matter halo around
NGC~2419 cannot be fully ruled out at present, yet any dark matter
within the 10\arcmin\ visible extent of the cluster must be highly concentrated and cannot exceed
$1.1\times 10^6\Msun$ (99\% confidence), in stark
contrast to expectations for a plausible progenitor halo of this structure.
\end{abstract}


\section{Introduction}
\label{sec:int}

Dwarf spheroidal galaxies (dSphs) and globular clusters (GCs) have
similar luminosities but are strikingly different in terms of their
dark-matter (DM) content, as inferred from the kinematics of their
stars: while dSphs (and their ``ultra-faint'' extension to low luminosity) 
galaxies are the most DM dominated systems, there is
virtually no evidence that GCs have DM halos
\citep[][]{Heg96,Mas05a,Bra11}.  This finding might represent a
problem for standard cold dark matter cosmology, in which GCs are
expected to form within their own DM halos \citep{Pee84}. However it
is quite possible that GCs were originally embedded in massive DM
halos that evolved during the cluster lifetime as a consequence of
interaction with the cluster stars \citep{Bau08} and stripping by the
tidal field of the host galaxy \citep{Mas05b,Sai06}. Therefore, the
central parts of GCs might be left relatively poor in DM, because the
DM at the present time either populates the outer region of the
cluster or has been stripped \citep{Bek12}.  However, this is unlikely to be the
case in GCs observed to have stellar tidal tails, which argue against
very extended massive DM components \citep{Moo96}.  While it is not
well established to what extent two-body relaxation and tidal
stripping are effective in shaping hypothetical DM halos of GCs, it is
clear that these effects are minimized in clusters with long two-body
relaxation time and that orbit the outer parts of the host galaxy,
where tidal effects are small. The prototype of these systems is the
remote massive GC NGC~2419, located in the halo of the Milky Way at a
distance of $d=87.5\kpc$ \citep{DiC11} (Galactocentric distance of $\sim 95 \kpc$), 
which has half-mass
relaxation time $\simeq 43\Gyr$ (\citealt{Har10} --- 2010 edition). If NGC~2419 formed
within a massive DM halo, it is reasonable to expect that such a halo
has maintained its properties almost unaltered up to the present time.

In this paper we use the recently published kinematic data set of 166
stars \citep[][hereafter Paper I]{Iba11a} and the analysis of the
stellar mass-to-light ratio $\Mstar/L$ \citep{Bel12} of the globular cluster
NGC~2419 to study the DM content of this system.  
Paper~I showed that the present kinematic and structural data
on NGC~2419 can be fit acceptably well with models containing no DM; 
but this does not necessarily imply that models with DM cannot be as good.
Previous studies
have tried to put upper limits on the DM mass of NGC~2419
\citep{Bau09,Con11}. \citet{Bau09} have argued that the observed
velocity dispersion profile of the cluster (based on 40 stars) is not
consistent with a DM halo more massive than $\Mdm=10^7\Msun$ (assuming
an isotropic stellar velocity distribution and NFW DM halo).
\citet{Con11} have put more stringent constraints ($\Mdm<10^6\Msun$),
based on the observed outermost light profile of the cluster. However,
the argument of \citet{Con11} relies on the hypothesis that the
cluster has relaxed via two-body relaxation, which is not thought to
be the case for NGC~2419.

Given the collisionless nature of NGC~2419 it is natural to explore,
besides models with isotropic velocity distributions, 
also models with radially or tangentially biased stellar orbits. 
Of course it is important to exclude very radially anisotropic systems, 
which are radial-orbit unstable. However, in
this respect it must be noted that stellar systems with massive DM
halos can sustain relatively more radial kinetic energy than
corresponding purely baryonic systems \citep{Nip02,Nip11}.

In Paper~I we have shown that it is very problematic to explain
the observed structure and kinematics of NGC~2419 if modified
Newtonian dynamics (MOND; \citealt{Mil83}) is the correct theory of
gravity (but see \citealt{San12} and \citealt{Iba11b}).  It must be stressed that this
result does not necessarily imply that NGC~2419 does not contain a
substantial amount of DM. In fact, what we can infer on the basis of
the \citet{Iba11a} results is that it is unlikely that NGC~2419 has a
MOND-equivalent DM halo, i.e. a DM halo with density distribution such
that the overall gravitational potential of the cluster is the same as
predicted by MOND \citep[see][and references therein]{Mil86,Nip11}. 
Of course, there is no reason to expect that the putative DM halo of
this globular cluster must mimic a MOND gravitational potential, so
it is quite possible that there is a significant amount of DM
in NGC~2419, but that this DM is not distributed as in a
MOND-equivalent halo\footnote{It must be noted that stellar systems with
MOND-equivalent halos can sustain more radial anisotropy in the
central regions than MOND stellar systems \citep{Nip11}, so, at least in
principle, it is also possible that the cluster is well represented by
very radially anisotropic models within a MOND-equivalent halo.}.

Throughout the paper we adopt the above distance to NGC~2419,
which implies an angular scale of
$25.452\pc$ per arcmin.  Our conclusions are virtually independent of
this specific choice, because the estimated uncertainty on the
distance is small ($\sim 4\%$; \citealt{DiC11}). We note that our
results on the dark-to-luminous mass ratio of NGC~2419, obtained for
given $d$ and stellar mass to light ratio ($\Mstar/L_V$), hold for all the combinations of $d$ and
$\Mstar/L_V$ such that $(\Mstar/L_V) \times d$ has the same value. In the present
work we account for the uncertainty on $\Mstar/L_V$, which is of the
order of $30\%$ \citep{Bel12}, so it is clear that our results are not
affected by the smaller uncertainties on $d$. Other physical
properties of the cluster are listed in Table~1 of Paper~I.

\section{Models without dark matter}
\label{sec:models_without_DM}

\subsection{Distribution function based models}
\label{sec:dynamical_models_noDM}

Before analysing the case for the presence of dark matter in NGC~2419,
it is instructive to consider models that possess only stars. 
In Paper~I (see also \citealt{Zoc12}), we 
presented several Michie model fits to the system in Newtonian gravity, and we 
showed that one of these models gives a statistically excellent representation of the 
cluster. However, due to high computational cost, only a small grid of 11 models 
were investigated in that contribution. To probe the parameter space much more finely,
we re-wrote our Michie model code to allow it to be run on a parallel computer,
thereby enabling us to calculate a large ensemble of Michie model fits, using a Markov-Chain Monte Carlo 
algorithm. This has the advantage of allowing us to 
properly incorporate the measurement uncertainties and also yields posterior probability
distribution functions for the fitted parameters. Furthermore, we are now able to use the 
stellar mass to light ratio measurement of \citet{Bel12} to constrain the solutions.

The integration of the Michie models (optionally residing within within a dark matter
potential) is outlined in Appendix A. 
We initially set the dark matter potential $\Wdm=0$ so as to suppress that component, leaving
5 free fitting parameters: the Michie model stellar central density, the core radius,
the anisotropy radius, the central potential and the stellar mass to
light ratio\footnote{In Paper~I, we analysed the effect on the line of sight 
velocity distribution of the presence of binaries in the cluster. The most likely binary fraction
was found to be modest ($\simlt 10\%$), so we decided to neglect this effect in the
present contribution.}. For a given set
of these 5 parameters, we construct a simulated surface brightness
profile and the line of sight velocity distribution as a function of
radius, which can be compared directly to the observations.

As in Paper~I, we convolve the model velocity distributions
projected onto the line of sight by the estimated Gaussian errors
associated to each kinematic datum $i$, to give a probability
distribution $f_p(v,R_i)$ (we drop the model superscript $j$ used in
our earlier contribution, since we now consider a continuum of models
defined by the above 5 parameters). The likelihood of a model then
becomes:
\begin{equation}
\ln L = \sum_{i} \ln[f_p(v_i,R_i)] - \sum_{k} {{(\Sigma_p - \Sigma_k)^2}\over{2 \delta\Sigma_k^2}} + \ln \Big[P\Big({{\Mstar}\over{L_V}}\Big)\Big]  \, ,
\label{eqn:likelihood}
\end{equation}
where the second summation over the ($k=15$) surface brightness
profile data points is included to take account of the likelihood of
the density profile, and the third term is the contribution from the
observed stellar mass to light $(\Mstar/L_V)$ ratio. 

In their analysis of recent WFC3 data of this cluster, \citet{Bel12}
discuss in detail why the stellar mass to light ratio
$(\Mstar/L_V)$ must be constant, independent of radius. This is deduced 
from the uniformity of the luminosity function throughout the cluster,
confirming an earlier analysis \citep{Dal08} that reached the same conclusions
based on the lack of radial segregation of the Blue Straggler Star (BSS)
population\footnote{The argument that justifies a constant stellar 
$M/L$ with radius is as follows: the BSS stars, which are 
significantly more massive than any evolving cluster star, are not observed
to be radially segregated in NGC~2419, i.e. they have the same radial
distribution as single stars. This should not be observed if mass
segregation is at work, hence 2-body relaxation is ineffective in this cluster.
Hence the stellar mix is the same at any radius, and therefore the stellar
mass-to-light is the same at any radius. This argument is valid
because mass segregation is expected (and observed)
in most classical globular clusters, where 2-body encounters 
are a relevant mechanism of energy-exchange (relaxation). In
less dense systems such as dwarf galaxies, the BSS population \citep{Map09} 
would not allow us to draw this conclusion, as they are fully non-collisional.}.
This justifies the simplicity of the second term on the right hand side of Equation~\ref{eqn:likelihood}:
the surface brightness profile should trace the projected mass density.

\begin{figure*}
\begin{center}
\vbox{
\hbox{
\includegraphics[bb= 20 20 360 350, angle=0, clip, width=8.5cm]{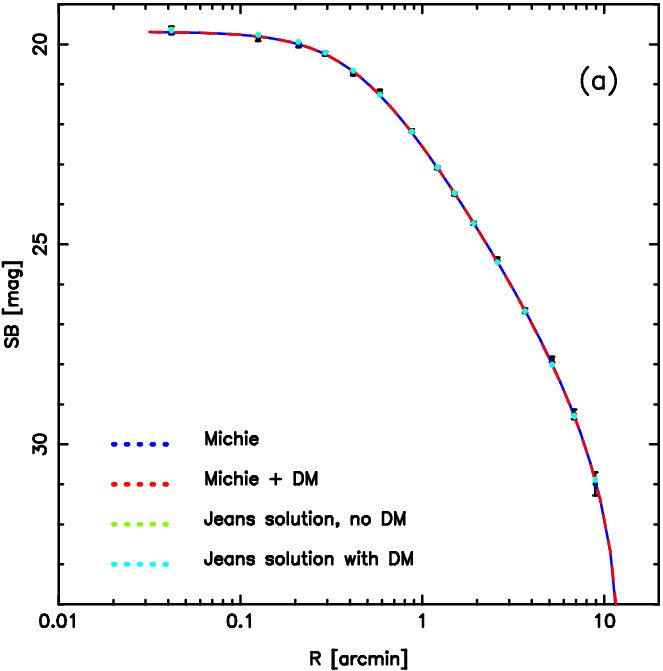}
\includegraphics[bb= 20 20 360 350, angle=0, clip, width=8.5cm]{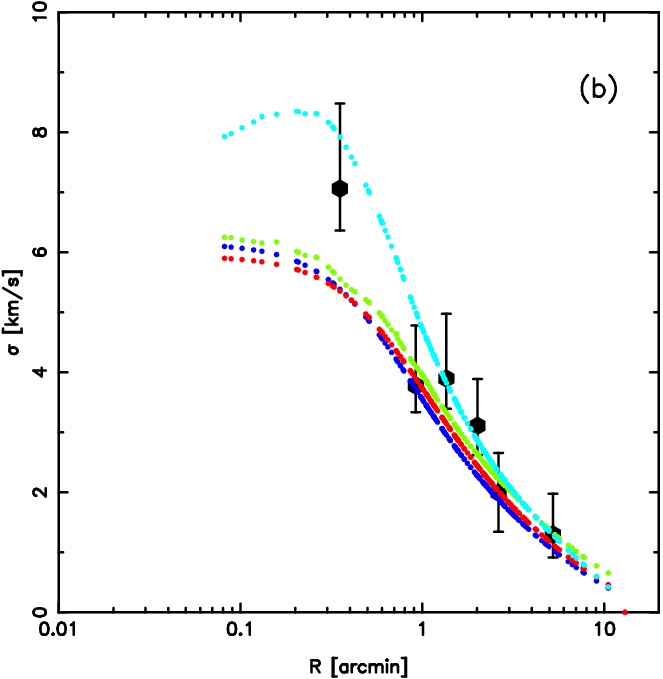}}
\hbox{
\includegraphics[bb= 20 20 360 350, angle=0, clip, width=8.5cm]{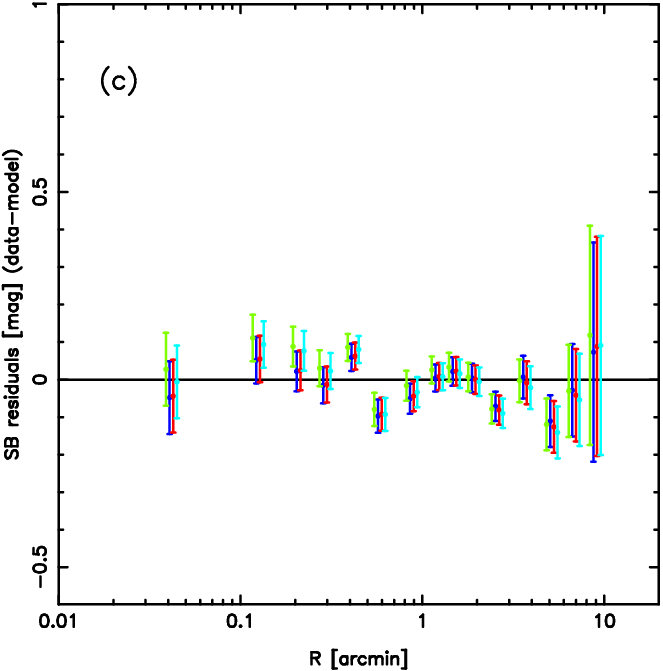}
\includegraphics[bb= 20 20 360 350, angle=0, clip, width=8.5cm]{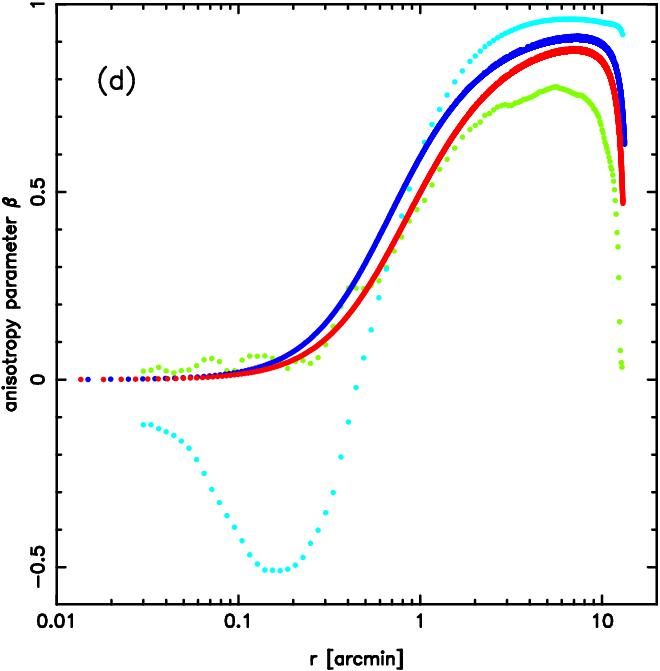}}
}
\end{center}
\caption{The surface brightness profile (a) and velocity dispersion profile (b) of the 
best-fit models. As described in the
text, the observed data were first fit with a pure Michie model (5 free parameters) using
an MCMC algorithm that explored this parameter space. 
The blue line in (a) shows the resulting
excellent fit to the surface brightness profile. The model
simultaneously also fits the kinematics, which are also shown in blue on the
right-hand panel. The big hexagonal markers show a binned
representation of the kinematic data, presented here {\em only} for
visual purposes. The likelihood calculation does not use these
binned values, but rather the individual kinematic measurements, the
radial locations of which correspond to the radial locations of the
small dots. The corresponding velocity dispersion of the model
provides the ordinate value of these points.
The best two-component model consisting of a Michie 
model to represent the stellar component and a double
power law model to describe the dark matter is displayed in red. These values
are almost coincident with the single (pure stellar) Michie model solution.
For this composite model, a total of nine free parameters were fitted by
the MCMC algorithm. The green and turquoise dots show the best non-parametric 
solutions derived with the MCMC Jeans equation solver. The model in green
has no DM, while the turquoise model allows for a DM component.
Panel (c): the surface brightness residuals between the data and the models
 (a small shift in radius has been applied to avoid the superposition of the
points). Panel (d) shows the $\beta$ anisotropy profile of the models.
The best non-parametric model without DM (green) was derived from $1.2\times10^7$ MCMC trials, 
probing a vast parameter space of solutions; the fact its $\beta$ profiles is very similar
to that of the pure Michie model (blue) lends strong support to the use of the Michie 
model as a good representation of this globular cluster.}
\label{fig:Michie}
\end{figure*}

Having fixed the shape of the density
profile, we allow for variation of its normalisation (and that of the
total stellar mass $\Mstar$) by leaving 
$\Mstar/\LV$ as a free parameter. \citet{Bel12} found a best-fit value of
$\Mstar/L_V=1.5\pm 0.1$, but they argue that uncertainty in stellar
evolution models allows for a range of $1.2 \la \Mstar/L_V \la 1.7$ in the
most likely value. For the probability density function $P$, we therefore
adopt a flat distribution between $1.2 \le \Mstar/L_V \le 1.7$, tapered as
a Gaussian of sigma 0.1 beyond these edges.

We investigated this model parameter space with a Markov-Chain Monte
Carlo (MCMC) approach, using the Parallel Tempering technique
\citep{Gre05}. The Parallel Tempering procedure explores the parameter
space at different ``temperature'' (i.e. smoothing) levels, which
ensures that the solutions do not get stuck in local maxima. In our
implementation we used 4 parallel MCMC chains, allowing swaps between
the chains at random intervals of approximately 100 steps (see
\citealt{Gre05} for a detailed description of this method). To ensure rapid 
convergence, the proposal step sizes of all the parameters together were 
refined (in lock-step) every 1000 iterations, while every 10000 steps the proposal step size 
of all the parameters were refined separately. We set a target acceptance ratio of 25\%,
so that on average one in four trial steps was accepted (this is considered to be optimal
when fitting a large number of parameters, see \citealt{Gre05}).

It is well known that systems with strong radial anisotropy
are prone to the radial-orbit instability.  Typically, stability
criteria are expressed in terms of the Fridman-Polyachenko-Shukhman
parameter $\xi \equiv 2 \Trad/\Ttan$, where $\Trad$ and $\Ttan \equiv
\Ttheta+\Tphi$ are the radial and tangential components of the kinetic
energy tensor, respectively \citep[][]{Pol81,Fri84}.  Here we find it
convenient to use the parameter $\xihalf$ which measures the same
ratio, but within the half-mass radius \citep{Nip11}.  In
paper~I we found that, in the absence of DM, $\xihalf=1.5$ can
be taken to separate stable and unstable models of NGC~2419. This
condition is used as a prior to reject unstable solutions.

The resulting best fit from 100000 iterations of the lowest temperature chain is
shown with the blue line and blue dots in Fig.~\ref{fig:Michie}, and is compared to the observational 
data. The corresponding Michie model parameter values and their uncertainties are listed in Table~\ref{tab:Michie}.
The solution is fully compatible with the best solutions found in Paper~I and \citet{Iba11b}.

\subsection{Jeans equation based models}
\label{sec:kinematic_models_noDM}

While the Michie model gives a statistically very good description of the cluster (as we have shown in Paper~I), 
one is nevertheless left wondering how unique this solution is. In that earlier contribution we showed that 
isotropic Michie models (commonly referred to as King models) cannot account for the 
observations, but one could attempt at this stage to fit alternative models such as a Plummer, a polytrope or some other model.
However, Nature need not follow these simple analytic forms.
To attempt to survey the vast space of more general solutions, we have developed an
algorithm that takes a different approach: we will assess instead the statistical properties
of non-parametric solutions to the Jeans equation.

For a spherically-symmetric system the Jeans equation \citep{Bin08}:
\begin{equation}
{{G M(r)}\over{r }} = - \sigma_r^2 \Bigg[ {{\mathrm{d} \ln \rho}\over{\mathrm{d} \ln
        r}} + {{\mathrm{d} \ln \sigma_r^2}\over{\mathrm{d} \ln r}} + 2\Big(1 -
    {{\sgth^2}\over{\sigma_r^2}} \Big) \Bigg] \, ,
\label{eqn:Jeans}
\end{equation}
relates $M(r)$ the cumulative mass inside a radius $r$ to the square of the velocity dispersion profiles
in the radial ($\sigma_r^2$) and tangential ($\sgth^2$) directions and to the 
logarithmic derivatives of $\sigma_r^2$ and of the density ($\rho$).

The algorithm we have developed explores these functions from the cluster centre out to the tidal radius.
To sample the functions in an economical manner we found it convenient to assign one third of the radial
bins to probe the inner region within two core radii with logarithmic bins. Beyond two core radii the 
remainder of the radial points used linear intervals. We set the value of the central velocity dispersion
in the radial direction to be a model parameter. The radial 
velocity dispersion profile is then defined by the sequence $\Delta \ln \sigma_r^2 (r_i)$, and the density profile is defined
by the sequence $\Delta \ln \rho (r_i)$, where $r_i$ are the locations of the radial bins. To interpolate these profiles into
logarithmic derivatives, we assume that the functions can be approximated locally (between adjacent radial 
bins) by parabolas in $\ln \sigma_r^2$ (or $\ln \rho$) versus $\ln r$. It is worth noting that it is
highly undesirable to define the profiles directly in $\sigma_r^2$ and $\rho$, since this requires
extremely high accuracy to avoid huge interpolation errors in the logarithmic derivatives.
The remaining parameters for the 
dark matter free case, are the total cluster mass $M_*$ and the stellar mass to light ratio $\Mstar/L_V$.

Our model therefore has 3 parameters unrelated to the chosen binning ($\sigma_r^2(0)$, $M_*$, $\Mstar/L_V$), plus
$(n-1)$ parameters that define the dispersion profile and $n$ that define the density profile. 
We found it convenient to set the number of radial bins to $n=1+2^m$, and certainly with $n=129$ the profiles
are smooth and we could reconstruct a Michie model very accurately from artificial data. 

The algorithm projects the 3-dimensional functions onto the line of sight, to produce a velocity dispersion profile
and a surface brightness profile, from which the model likelihoods are measured using the observed data 
(Eqn.~\ref{eqn:likelihood}). 
One further assumption is made in the calculation of the likelihood: we assume that the velocity distributions
are Gaussian. This assumption is not necessary for the algorithm to work, and we could have used
the velocity moments of the kinematic data directly, or indeed we could have probed other parametrisations
of the velocity distribution. However, it is computationally faster and avoids the errors deriving 
from the calculation of the moments from the data. Furthermore, 
the assumption is supported by the data: as we showed in 
Paper~I, the observed distribution is consistent with a Gaussian function at all radii probed.
The line of sight velocity distributions are then convolved with the uncertainty distribution appropriate for each data
point.

We explore this model ($260$ free parameters) using a MCMC 
scheme\footnote{The Bayesian method we are using yields probability distribution functions
for the fitting parameters, as well as information on the correlations between these parameters. In this context it
is not a problem that we have more fitting parameters than data points.}. As before, we use the Parallel
Tempering technique with 4 chains to probe progressively more smoothed versions of the likelihood surface.
Following \citet{Gre05}, this is achieved by multiplying $\ln(L)$ by a factor $1/m^2$, where $m$ is the chain number.
The $m=1$ chain corresponds to the lowest ``temperature'' simulation, which is the one of primary interest. 

We implemented an improvement on this method, reducing the dimensionality of the high temperature
chains by redefining their radial profiles at only $n=1+2^m$ points. Cubic interpolation (in $\ln \sigma_r^2$ and $\ln \rho$ 
versus $\ln r$) between these anchor points was used to fill in intermediate values to maintain 129 radial 
bins in all chains. Although we attempted a simulation including a $m=5$ chain (i.e. with 9 radial anchor points), 
it was clear that the interpolation was too crude to be useful.

Some regularisation is required to tell the algorithm that it should avoid exploring very irregular profiles. This was implemented
by using a mild smoothing prior on the profiles of the logarithmic derivatives.

An initial simulation with $2\times10^6$ iterations for each chain was run using a Metropolis-Hastings
acceptance criterion. As before, all parameters were tuned in lock-step every 1000 iterations, and then 
refined independently every $5\times10^5$ iterations, so as to ensure an acceptance criterion of 25\%.
Finally, the 1000 best solutions from this Metropolis-Hastings algorithm were used as inputs for another
$2\times10^6$ iteration run using an affine-invariant ensemble sampler algorithm. Our algorithm
is based on the ``stretch move'' algorithm proposed by \citet{Goo10}, which we expand to use the 
above-described multiple ``temperature'' parallel chains. The simulation uses a population of 1000
``walkers'' (which are non-coincident parameter space positions) which evolve from one iteration
to the next, and so sample efficiently the likelihood function.
The convergence of the method was checked by running simulations from 3 very different starting positions.

While the models built from the distribution function (described in
Section~\ref{sec:dynamical_models_noDM}) are consistent by construction, 
this Jeans-equation based approach can yield solutions
that do not have physically possible distribution functions. This is a
fundamental drawback of the technique, and cannot be avoided. 
However, some unphysical models are eliminated by ensuring that the solutions
obey the Global Density Slope-Anisotropy Inequality \citep[GDSAI;][]{Cio10a,Cio10b},
which is the requirement that at each radius the negative of the
logarithmic slope of the stellar density distribution
$\gamma_*\equiv -\d\ln \rhostar/\d\ln r$ cannot be less than twice the
local value of the anisotropy parameter, $\gamma_*(r)\geq 2\beta(r)$.
We remind the reader that the anisotropy parameter is defined as:
\begin{equation}
\beta(r) \equiv 1 -{\sgthsq+\sgphsq\over 2\sgrsq},
\end{equation}
where $\sgr$, $\sgth$ and $\sgph$ are, respectively, the $r$,
$\vartheta$ and $\varphi$ components of the velocity-dispersion
tensor. As in Section~\ref{sec:dynamical_models_noDM}, we also
impose the requirement that $\xi_{half} < 1.5$ to ensure stability of the resulting models.

In Fig.~\ref{fig:Michie} we display, in green, the surface brightness profile and 
velocity dispersion profile of the best solution to the Jeans equation found with this 
algorithm. The anisotropy profile of this solution is also shown (panel d); 
it is striking how closely this non-parametric solution matches
the behaviour of the best-fit Michie model from Section~\ref{sec:dynamical_models_noDM}
(shown in blue). 
That our Jeans equation solving algorithm should have selected this model, which is
so similar to the previously-fitted Michie model, out
of the vast space of possible models, lends strong support to the use of the Michie distribution function
to represent the stellar component in this system.
The small differences with respect to the Michie model fit in Fig.~\ref{fig:Michie} can be
attributed to the differences in the anisotropy profile but the
driving factor is the different stellar mass to light ratio (${\Mstar/L_V}=1.89$).

\section{Models with dark matter}
\label{sec:models_with_DM}

Following the layout of the previous section, we consider first adding dark matter
to the Michie model, and then freeing $M(r)$ in the Jeans
equation so as to also allow for dark matter in those solutions.

\subsection{Distribution function based models}
\label{sec:dynamical_models_DM}

We first consider the possibility that the putative dark-matter halo can be described by the double power law density profile
\citep{Bin08}:
\begin{equation}
\rhodm (r)=\frac{\rhodmzero}{\left(\frac{r}{\rs}\right)^\gamma\left(1+\frac{r}{\rs}\right)^{\delta-\gamma}},
\label{eq:twopower}
\end{equation}
where $\gamma$ and $\delta$ are the two power-law indices
characterising the shape of the profile, $\rs$ is a scale radius, and
$\rhodmzero$ is a reference density.  In particular, when $\gamma=1$
and $\delta=3$ we get the \citet[][NFW]{Nav96} model; when $\gamma=0$
and $\delta=3$ we get a profile with a flat inner density
distribution, which is very similar to the profile proposed by
\citet{Bur95}. For combinations of $\gamma$ and $\delta$ such as those
mentioned above the integrated mass of the model diverges at large
radii, so it is useful to introduce as an additional parameter a
truncation radius $\rvir$ such that $\rhodm=0$ for $r> \rvir$.  The DM
mass contained within a radius $r$ is
\begin{equation}
\Mdm(r)=4\pi \int_0^{r}\rhodm(r')r'^2dr',
\end{equation}
so the total mass of the DM halo is $\Mdmvir=\Mdm(\rvir)$.  If the
stellar density distribution is truncated at $\rt$, the kinematics of
the cluster is influenced only by the mass distribution within 
the stellar tidal radius $\rt$.
It follows that our results are not sensitive to the value of $\rvir$,
provided that $\rvir\geq\rt$, which we assume throughout the paper (in
other words, our results are not sensitive to the halo concentration
$C\equiv\rvir/\rs$). As a consequence, we define the quantity
$\Mdmt\equiv\Mdm(\rt)$ (the DM mass contained within $\rt$), which is
more useful than $\Mdmvir$ in order to characterise our models.  In
summary, our model halo is determined by specifying the following four
parameters: $\gamma$, $\delta$, $\rs$ and $\Mdmt$.

For the DM component, for any choice of the DM profile parameters ($\gamma$ and $\delta$)
the adimensional potential can be calculated by direct integration of equation~(\ref{eq_pois}) as
$$\Wdm(r)=9~
\frac{\rhodmzero}{\rho_{*,0}}\left(\frac{\rs}{\rc}\right)^{2}\int_{r/\rs}^{\rt/\rs}\frac{1}{r'^{2}}\int_{0}^{r'}r''^{2}\rhodm(r'')dr''~dr'$$
In the particular cases of a NFW ($\gamma=1;~\delta=3$) and a cored
($\gamma=0;~\delta=3$) halo the above integral admits the analytical solutions:
\begin{equation}
\label{notev_eq}
\Wdm(r)=9~\frac{\rhodmzero}{\rhostarzero}\left(\frac{\rs}{\rc}\right)^{2}~\frac{\ln\left(1+\frac{r}{\rs}\right)}{\frac{r}{\rs}} 
\end{equation}
when $(\gamma=1,\delta=3)$, and
\begin{equation}
\Wdm(r)=-\frac{9}{2}~\frac{\rhodmzero}{\rhostarzero}\left(\frac{\rs}{\rc}\right)^{2}~\frac{\frac{r}{\rs}-2\left(1+\frac{r}{\rs}\right)~\ln\left(1+\frac{r}{\rs}\right)}{\frac{r}{\rs}\left(1+\frac{r}{\rs}\right)}
\end{equation}
when $(\gamma=0,\delta=3)$.  

By incorporating this dark matter potential into the procedure
described in \S\ref{sec:dynamical_models_noDM}, one
can construct spherically symmetric
self-consistent stellar systems, truncated at a tidal radius
$\rttilde$, in equilibrium in the presence of a DM halo \citep[see
  also][]{Amo11}. In these models the parameters $\Wzero$, $\rs/\rc$,
$\ratilde$, $\rhodmzero/\rhostarzero$, $\gamma$ and $\delta$ drive the
shape of the density and velocity dispersion profiles, while the
values of $\rhostarzero$ and $\rc$ determine their normalisation to
physical units (see Appendix).  

The stellar components of the models considered here are self-consistent by
construction, but in principle it is not guaranteed that there is a
positive distribution function generating the DM components. For
a wide class of two-component models, a necessary condition for
consistency is that also the DM component satisfies the GDSAI \citep{Cio10b}.
Clearly the specific form of the halo anisotropy
profile is not important for the purpose of the present investigation,
because the halo influences the observables only through its
gravitational potential. However, we impose that in our model the halo
density never increases with radius ($\d\ln \rhodm/\d\ln r\leq0$ at
all $r$, i.e. $\gamma,\delta > 0$). This is a necessary condition for consistency only when the
halo is isotropic or radially anisotropic, but we impose it as a
general constraint because we think it is an astrophysically motivated
assumption. However, we noted that it is not excluded that
there are self-consistent tangentially biased models with
$\d\ln\rhodm/\d\ln r<0$ at some $r$.

Even if consistent, some models can be excluded because they are
unstable. It has been shown that DM halo has a stabilising effect against the
radial-orbit instability \citep{Nip02,Nip11}, so at least some models
with DM halo are expected to be stable even if they have
$\xihalf>1.5$.  However, a full stability analysis of two-component
models of NGC~2419 is beyond the purpose of the present paper, so here
we conservatively assume $\xihalf\leq 1.5$ as a stability criterion
for all the models considered in the present work.  With this choice
we should effectively exclude unstable models.

We proceed to examine the relative likelihoods of the
two-component models built from a power-law dark matter halo and a
Michie model stellar component. As in \S\ref{sec:dynamical_models_noDM}, we
use a ``Parallel Tempering'' MCMC algorithm with 4 chains to explore
the parameter space. The composite model has 9 free
parameters: the stellar central density, the Michie model core radius,
the anisotropy radius, the central potential, the stellar mass to
light ratio, the dark matter central density, the DM characteristic
radius and the DM inner and outer power-law indices. For a given set
of these 9 parameters, we construct a simulated surface brightness
profile and the line of sight velocity distribution as a function of
radius, which can be compared directly to the observations.

Several (very weak) priors were imposed on the solutions. The stellar core radius
must lie within a factor of 10 of $10.88\pc$; the anisotropy radius
has to lie between 0.5 and 100 core radii;
the central potential parameter has to lie between 2 and 20; and the
stellar mass to light ratio is required to stay within the bounds of
$0.8<{\Mstar/L_V}<2.1$ (i.e. within $4\sigma$ of the flat region 
$1.2<{\Mstar/L_V}<1.7$ stated above in Section~\ref{sec:dynamical_models_noDM}). 
Furthermore, we impose that the characteristic
radius of the dark matter has to lie between $25\pc$ and $2.5\kpc$;
and that the inner and outer power law exponents lie in the ranges $0<
\gamma< 1.5$ and $2<\delta<4$. While certain black hole growth models
suggest that the inner power-law slope may be higher than $\gamma = 1.5$
(e.g., \citealt{Qui95}), this is likely to affect only the very central region of the
cluster where our data do not constrain the density profile.

\begin{figure}
\begin{center}
\includegraphics[bb= 30 20 570 570, angle=0, clip, width=8.5cm]{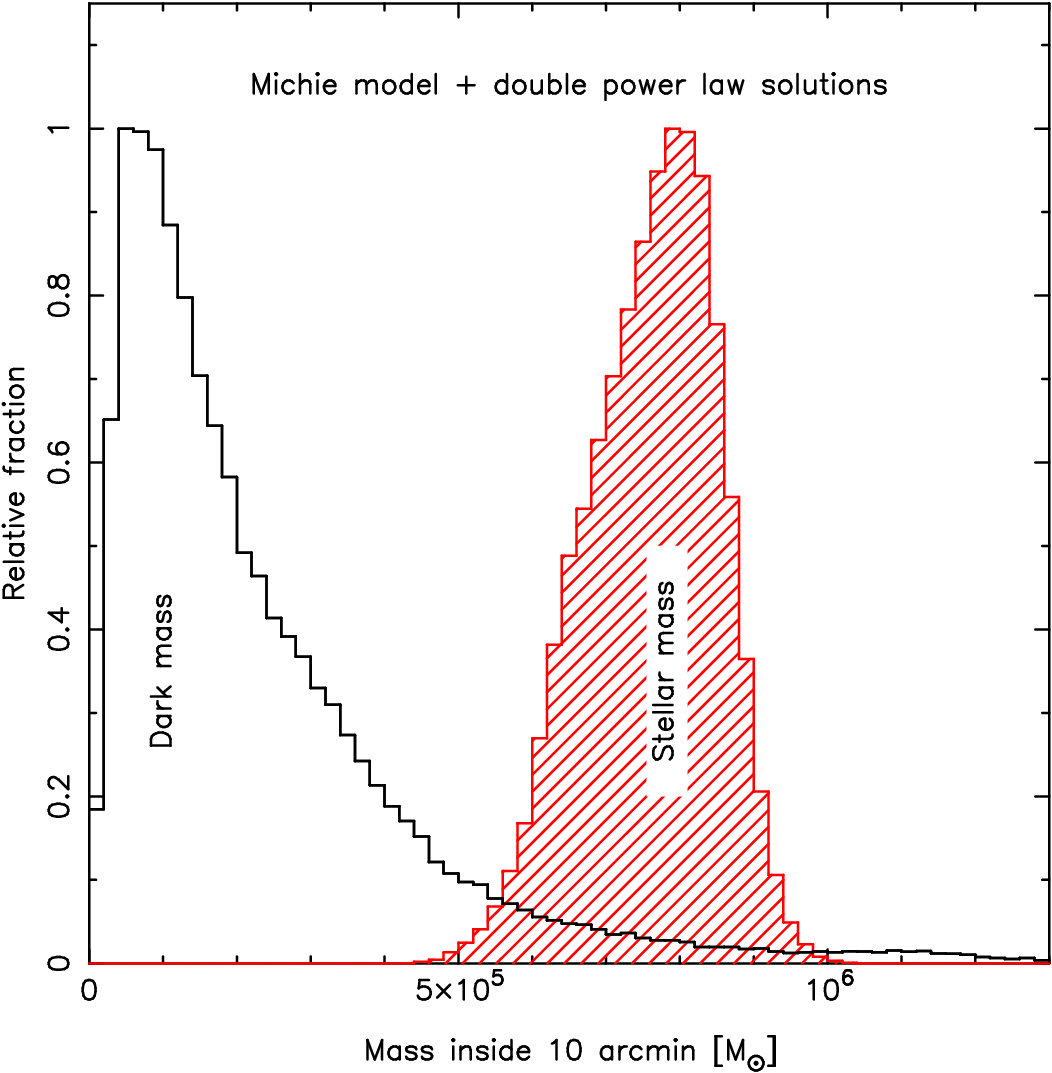}
\end{center}
\caption{Unshaded histogram: the posterior distribution of the mass of the dark component
  inside of $10\arcmin$ (the approximate size of the stellar
  structure) derived by the MCMC algorithm that fits a Michie model
  inside a dark matter halo modelled as a double power-law. The hatched
  histogram shows the posterior distribution of the luminous component 
  of the cluster. Thus if we impose a
  Michie model on the luminous component (which as we have shown in
  Paper I, gives an excellent fit to the observations), the most
  likely solutions for the dark matter have very little mass relative
  to the stars. It would be natural to conclude that the cluster needs
  no dark component.}
\label{fig:composite_posterior}
\end{figure}

Figure~\ref{fig:Michie} shows (in red) the most likely model found by the
MCMC algorithm in $10^6$ iterations. The solution has $\rc=11.4\pc$,
$\ra=1.6\rc$, a dark matter characteristic radius of $28\pc$, and
$\gamma=0.06$ and $\delta=3.7$. However, we were surprised to find that
this model requires very little dark matter:
$\Mdm=5.0\times10^4\Msun$ inside of 10\arcmin\ (the approximate limit
of the stellar structure), which is effectively insignificant compared
to the stellar mass of $8.1\times10^5\Msun$. Marginalising over all
other parameters, we show the posterior distribution of $\Mdm$ in
Fig.~\ref{fig:composite_posterior}; clearly the data appear to favour
models without substantial dark matter.

To confirm this curious result, we re-ran the MCMC experiments forcing
certain astrophysically-motivated parameter sets: cored DM models
($\gamma=0$) and the NFW model ($\gamma=1$, $\delta=3$). In an almost
identical manner to the more general solution above, the model
likelihoods are found to decrease strongly with increasing DM
fraction. In an attempt to explore whether the MCMC algorithm was able
to escape from solutions with low dark matter, we also performed an
experiment forcing a massive DM component, by setting $\Mdm>
5\times10^6\Msun$ as a strong prior. But again, the distribution of
solutions simply clustered near this low mass limit.

We conclude from this analysis that a simple Michie model where mass
traces light is actually preferred by the data over more complicated
models that include the possibility of having dark matter in the form
of a double power-law profile. The preference for a cored distribution
turns out to be very slight - the best models with $0.9 < \gamma < 1.1$
are only a factor of 1.4 less likely than the best model.
However, solutions with some DM seem better than those without DM. 
This is a manifestation of the tension between stellar and dynamical 
mass to light ratio that was noted in \citet{Bel12} (and also in \citealt{Sol12} 
in other clusters) and which is discussed further below.

\begin{figure}
\begin{center}
\includegraphics[bb= 30 20 570 570, angle=0, clip, width=8.5cm]{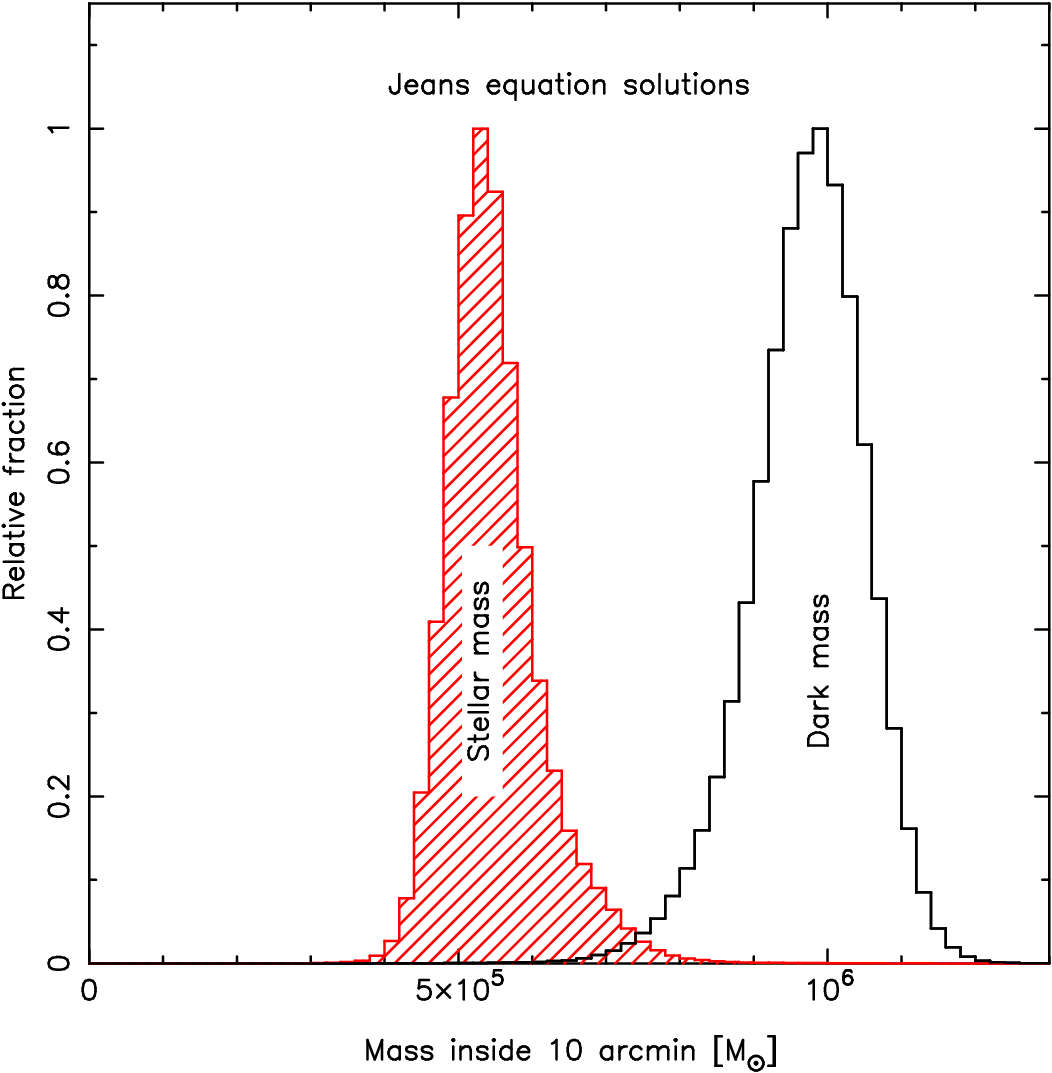}
\end{center}
\caption{Unshaded histogram: the posterior distribution of the mass of the dark component
  inside of $10\arcmin$, but this time calculated from an MCMC
  algorithm that solves the Jeans equation in a non-parametric way.
  A total of $1.2 \times 10^7$ MCMC iterations are presented. 
  The difference to Figure~\ref{fig:composite_posterior} is striking:
  with the freedom from analytic form, the preferred solutions
  possess a non-negligible dark matter component. The
  stellar mass of the models (hatched histogram) is approximately half 
  that of the dark matter.}
\label{fig:Jeans_posterior}
\end{figure}

\subsection{Jeans equation based models}
\label{sec:kinematic_models_DM}

The paucity of dark matter implied by our MCMC survey of the likelihood
surface of Michie models is compelling, since the Michie models form
a fairly general family of structures, but one is again left wondering to what 
extent this conclusion depends on the choices of the models
selected to examine both the cluster and the dark matter. 

For this reason we extended our analysis to a
more general set of spherical models where both the dark matter density
profile and the stellar density profile are free non-parametric profiles. 
We reuse the non-parametric MCMC tool
described in \S\ref{sec:kinematic_models_noDM} which solves the Jeans 
equation, finding solutions that are consistent with the data and our priors.

\begin{figure*}
\begin{center}
\hbox{
\includegraphics[bb= 30 20 570 570, angle=0, clip, width=8.5cm]{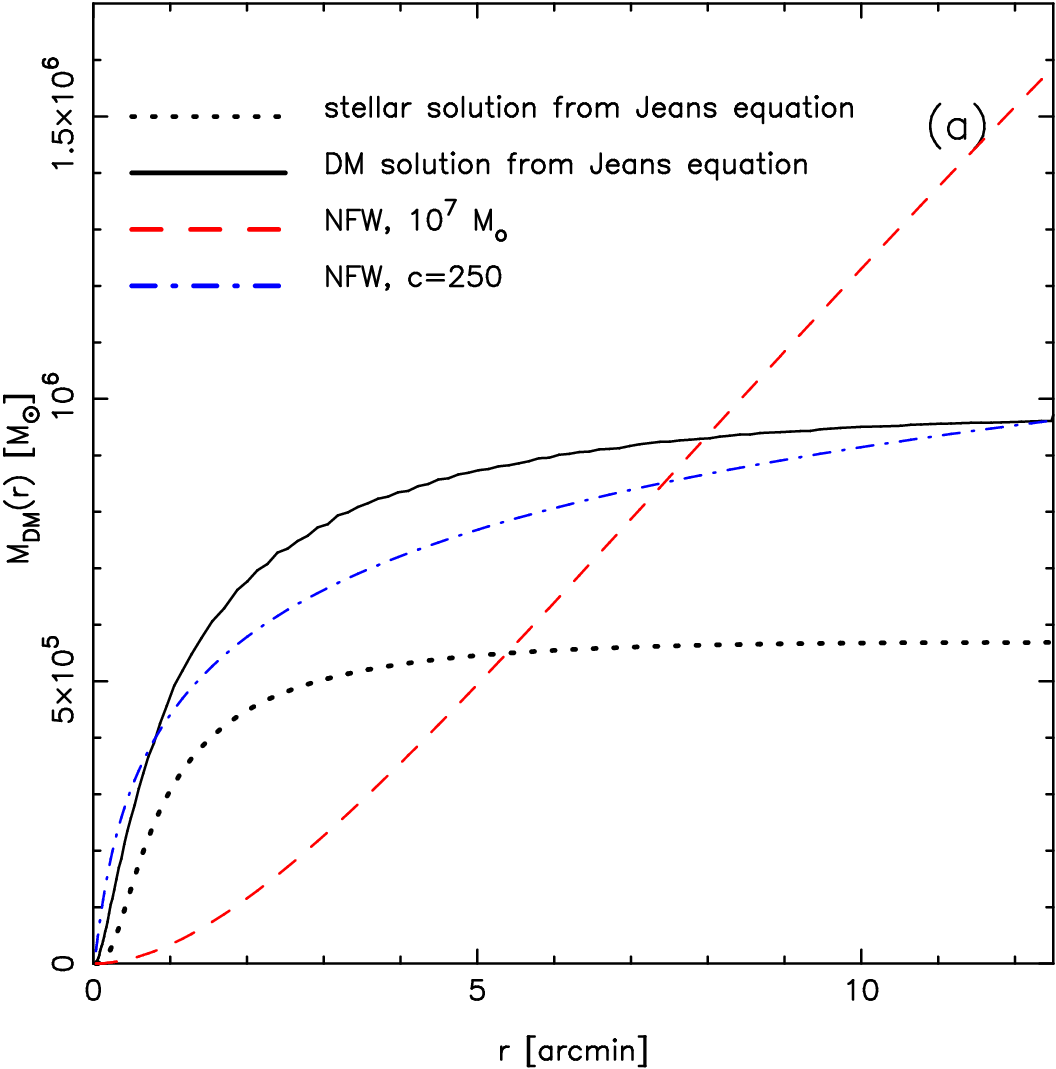}
\includegraphics[bb= 30 20 570 570, angle=0, clip, width=8.5cm]{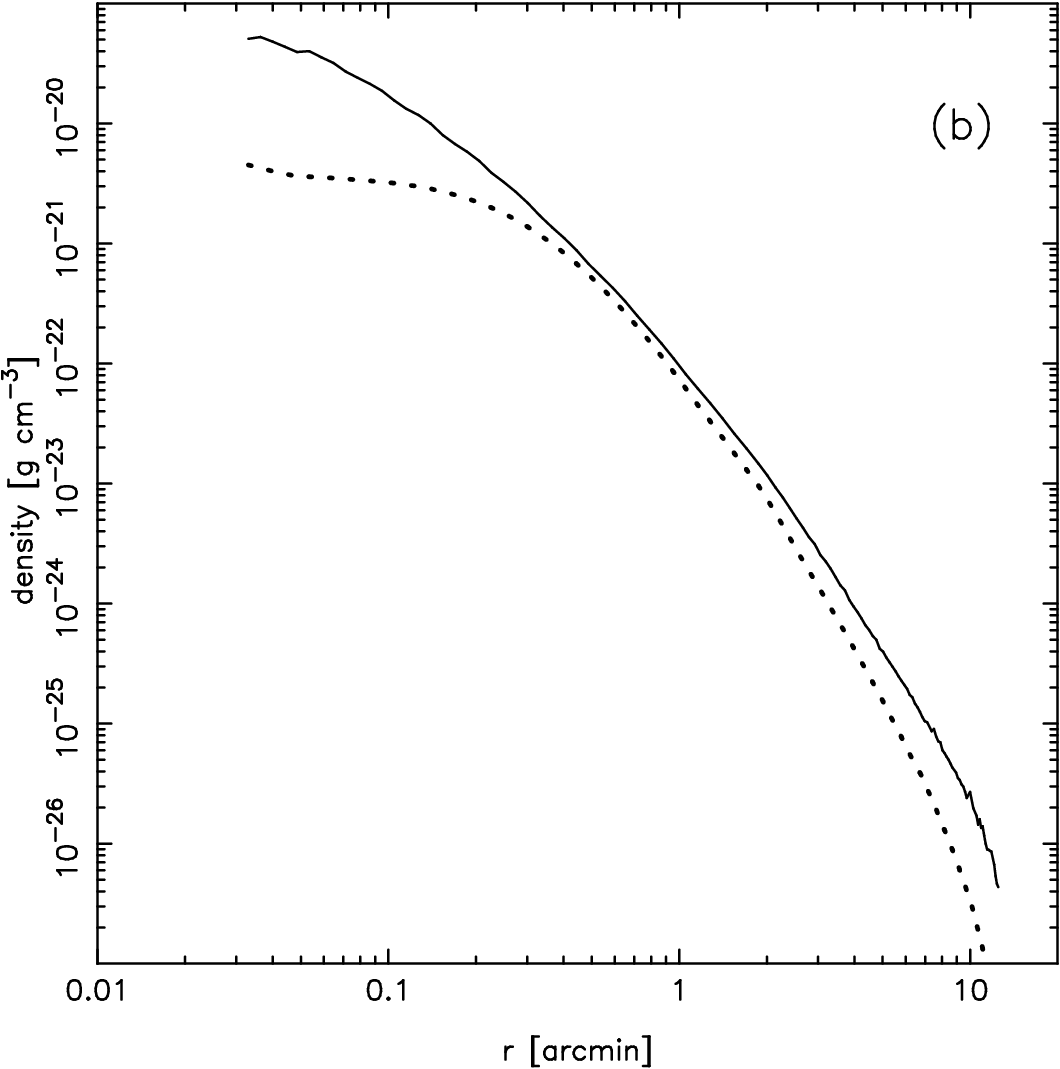}
}
\end{center}
\caption{Panel (a): The cumulative mass profile of the dark matter component of
  the most likely model derived by the MCMC algorithm using the non-parametric Jeans
  equation method is shown with the continuous line. The dotted line
  shows the cumulative mass profile of the stellar component of the same solution.
  To first approximation the dark matter profile appears fairly similar to the stellar
  profile, and stands in stark contrast to cosmological expectations: the dashed line shows
  an NFW profile with $M_{vir}=10^7\Msun$, concentration $c=23.5$ and $r_{vir}=4.5\kpc$. The dot-dashed 
  line shows an attempt to mimic the continuous line with an NFW profile; this however
  requires an extremely high concentration ($c=250$), even if $r_{vir}=13\arcmin = 0.33\kpc$.
  Panel (b): the corresponding density profiles.}
\label{fig:mass_profile}
\end{figure*}

However, given that we now allow for the possibility of dark matter, 
we require some additional information to close equation~\ref{eqn:Jeans}.
To this end we have chosen to free the tangential velocity dispersion profile (in addition
to $\sigma_r^2(0)$, $M_*$, ${\Mstar/L_V}$, and the profiles of 
$\Delta \ln \rho$ and $\Delta \ln \sigma_r^2$).
We found it convenient to tabulate this profile as $\Delta \ln \sgth^2$, adding
a further $n-1$ parameters to the MCMC fit, plus $\sgth^2(0)$. The model
thus has a total of 389 free parameters.

The MCMC simulations were run in an identical manner to those
described in \S\ref{sec:kinematic_models_DM}, using again 
a hierarchy of 4 different ``temperature'' chains, each with progressively lower
spatial resolution. As before, after initial runs with $2\times10^6$ 
iterations with a Metropolis-Hastings acceptance algorithm, we
switched to using a population of affine-invariant ensemble samplers and
reran the simulation for a further $2\times10^6$ iterations.

While the models built from the distribution function (described in
Section~\ref{sec:dynamical_models_DM}) are consistent by construction (at least
for as far as the stellar component is concerned), we cannot exclude that some of the  two-component models
considered are not generated by positive distribution functions. However, as in Section~\ref{sec:kinematic_models_noDM}
we impose at least some necessary conditions for consistency. We additionally
imposed the prior that the dark matter density profile decreases with radius
and that its logarithmic slope becomes increasingly negative with increasing radius, i.e. if
the radial locations $r_1$ and $r_2$ are such that $r_1 < r_2$, the logarithmic slopes
at these locations obey the condition:
$$ 0 > {{\d\ln \rhodm}\over{\d\ln r}} \Big|_{r_1} > {{\d\ln \rhodm}\over{\d\ln r}} \Big|_{r_2} \,.$$
This property is shared by many commonly-used models such as the Plummer sphere,
the Michie (and hence King) models, the double power law models if $\delta > \gamma$
(of which the NFW is a member), among others.

\begin{figure}
\begin{center}
\includegraphics[bb= 30 20 570 570, angle=0, clip, width=8.5cm]{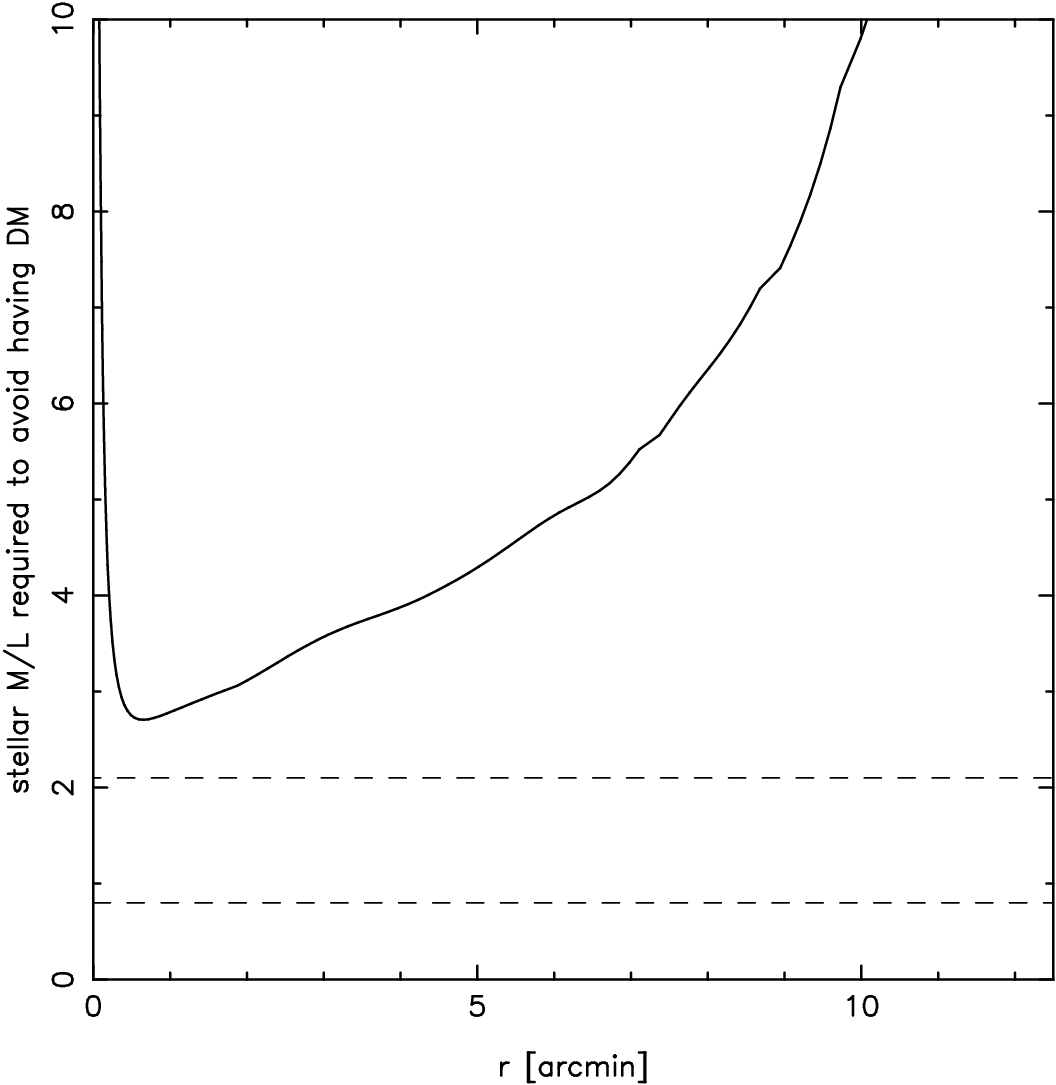}
\end{center}
\caption{For the most likely non-parametric solution to the Jeans equation 
(shown previously in Figure~\ref{fig:mass_profile}), we calculate the stellar 
mass to light ratio required to eliminate the need for a dark matter component. The region between
the two dashed lines marks the maximum ($4\sigma$) range consistent with the analysis of \citet{Bel12}.}
\label{fig:ML}
\end{figure}

\begin{figure*}
\begin{center}
\vbox{
\hbox{
\includegraphics[bb= 20 20 360 350, angle=0, clip, width=8.5cm]{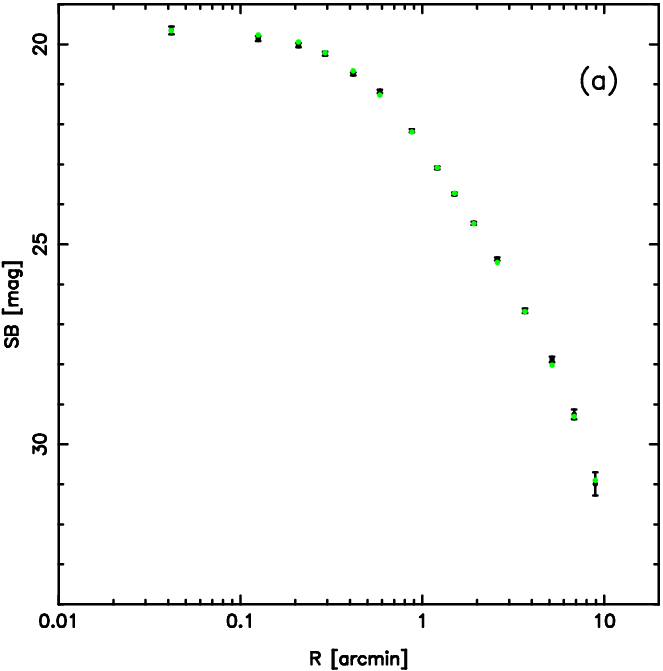}
\includegraphics[bb= 20 20 360 350, angle=0, clip, width=8.5cm]{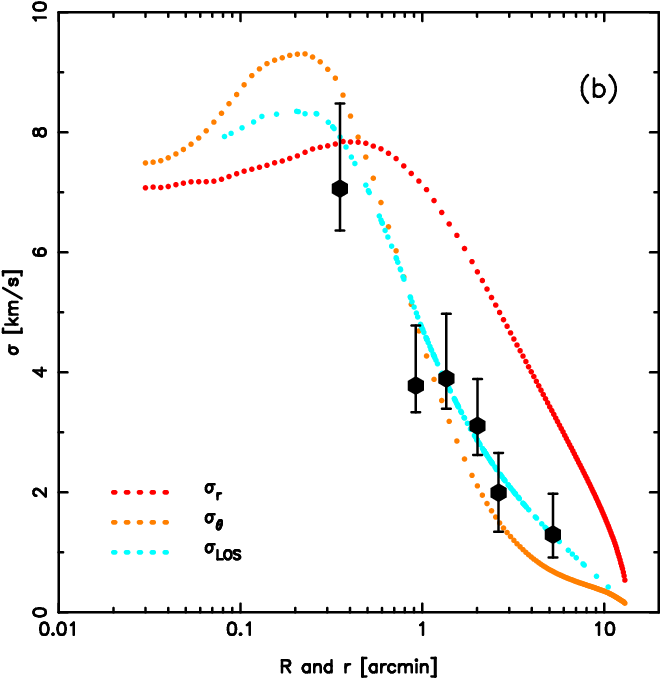}}
\hbox{
\includegraphics[bb= 20 20 360 350, angle=0, clip, width=8.5cm]{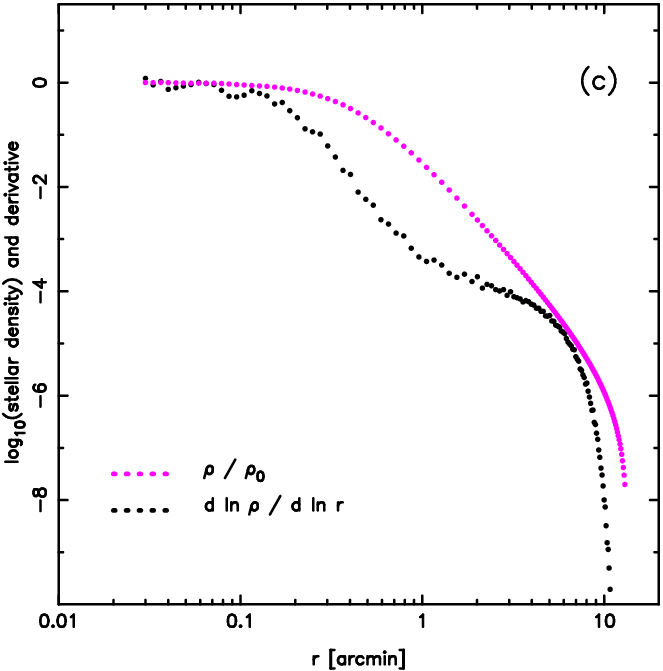}
\includegraphics[bb= 20 20 360 350, angle=0, clip, width=8.5cm]{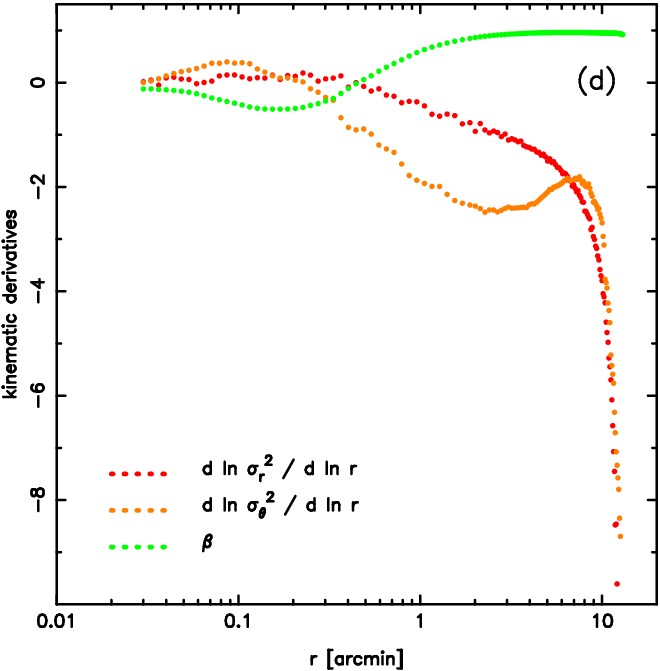}}
}
\end{center}
\caption{
The most likely Jeans equation solution of NGC~2419 using the new non-parametric MCMC technique
allowing for both a stellar and a dark matter component.
Panels (a) and (b) show the fit to the observed surface brightness and kinematics, respectively.
The data are reproduced from Figure~\ref{fig:Michie}. The small green dots in (a) display
the excellent model fit. The non-parametric velocity dispersion profiles in the radial and tangential
directions ($\sigma_r$ and $\sgth$) along with their line of sight projection ($\sigma_{LOS}$)
are displayed alongside the velocity dispersion estimates. We stress again that the {\em individual}
star velocities were used in the likelihood calculation. The 
remaining profiles required for the Jeans equation solver are displayed in the bottom panels: the {\it stellar}
density and its derivative are shown in (c), and the velocity dispersion derivatives in (d). 
For reference, we also show in panel (d) the anisotropy profile derived from the velocity 
dispersion profiles shown in (b). Note that the logarithmic derivative 
profiles are significantly noisier than the physical profiles. This model has a stellar mass to 
light ratio ${\Mstar/L_V}=1.20$, 
which is within the acceptable range reported by \citet{Bel12}. }
\label{fig:Jeans_composite}
\end{figure*}

Figure~\ref{fig:Jeans_posterior} shows the
marginalised posterior distribution of the dark matter in the solutions (unshaded 
histogram). The distribution peaks at $\approx 9.7\times 10^5 \Msun$, suggesting the
presence of a significant dark matter fraction within the central 10\arcmin. 
The posterior distribution of the mass of the stellar component, shown with the hatched histogram,
reveals that roughly half the amount in stars as dark matter appears to be present in the 
solutions. At first sight it would appear that we are facing a simple degeneracy between the total
mass of the cluster and the mass to light ratio of the stellar population. This suspicion 
is reinforced by Figure~\ref{fig:mass_profile}a, which shows the cumulative mass profile
in the most likely solution: the dark matter (solid line) and stars (dotted line) follow each other
fairly closely (but see also the density profile in panel b). 
This best solution has a stellar mass to light ratio ${\Mstar/L_V}=1.20$, 
within the acceptable
range found by \citet{Bel12}. This would therefore suggest that the ``dark matter'' component is simply 
the result of the algorithm attempting to force a constant stellar ${\Mstar/L_V}$ over the whole radial range of
the cluster.

However, this option does not appear to be possible, since it would require a ${\Mstar/L_V}$ profile that is
inconsistent with observations and other astrophysical priors, as we show in Figure~\ref{fig:ML}. 
Here we have calculated the stellar ${\Mstar/L_V}$ that would be required to account for the total density at 
a given radius.
The constraints on the uniformity of the radial profile of the ${\Mstar/L_V}$ \citep{Bel12}, as well as
the implausible ${\Mstar/L_V}$ values beyond $\sim 5\arcmin$ make it unlikely that the dark matter component in our
Jeans equation solutions are simply due to erroneous assumptions in the ${\Mstar/L_V}$ profile.

The best Jeans equation solution is displayed in Fig.~\ref{fig:Jeans_composite}; the panels 
also illustrate the working of the method by showing the non-parametric profiles 
that are refined at each Markov-Chain Monte Carlo
iteration. 
As can be seen, the solution gives an almost perfect reproduction of the surface brightness
data, and interestingly, is able to fit the full range of the kinematic data, giving a
higher (and better) central velocity dispersion than any of the Michie models studied here or
in Paper~I. Beyond the central $\sim 1\arcmin$, the model is highly radially 
anisotropic\footnote{Recently, \citet{Zoc12} have studied NGC~2419 (among other clusters)
using so-called $f^{(\nu)}$ models, which can represent violently relaxed systems. In their
analysis (which assumes that no dark matter is present), they also manage
to produce a steep central velocity dispersion profile (see their Fig.~3).  Note
however that their best-fit $f^{(\nu)}$ model has ${\Mstar/L_V}=2.4$ and is quite
anisotropic $(\xi=1.77)$.}.

\section{Discussion}
\label{sec:Discussion}

With current data the answer to the question of whether the globular cluster NGC~2419
contains dark matter or not depends on the model (and hence one's priors) used to analyse the structure.
The choice of using a Michie model for the stars plus a double power law dark matter halo,
is appropriate for those who value highly dynamically self-consistent methods and who adhere 
strongly to cosmological expectations. 
The most likely solution of this composite model had $M_{DM}(<10\arcmin)=5\times 10^4\Msun$, 
which would imply in standard
$\Lambda$CDM cosmology (following \citealt{Nav00})
a virial mass of $\sim 10^5\Msun$, roughly an order of magnitude lower than the
mass of the stellar component in the cluster. 
The upper 99\% confidence limit for the dark mass inside of $10\arcmin$
according to the probability distribution function in Fig.~\ref{fig:composite_posterior} is $7.2\times 10^5\Msun$, 
and this corresponds to a virial mass of $\sim 4\times 10^6\Msun$.
Such a minuscule halo is some two orders of magnitude lower than expectations for the halo mass of dwarf 
spheroidal galaxies \citep{Wal12}, which have similar stellar content. 
Thus with this composite model the data prefers virtually no dark matter and one can rule out 
the possibility that the cluster is embedded at the centre of a dwarf-spheroidal-like mini-halo.

The issue of whether the cluster is located at the centre of the putative DM halo is important
for the validity of the analysis presented here. However, the case should in principle be rather similar
to that of the globular cluster M54 in the Sagittarius dSph, where the dynamical friction 
timescale for this similarly-massive globular cluster and its host halo to coincide is estimated to 
be $\sim 1.5\Gyr$ \citep{Bel08}. The expectation therefore is that NGC~2419 should 
reside at the DM halo centre, if it possesses one.

The alternative analysis, using our new Jeans equation solver yields a slightly
different answer. The strength of this method is that it does not require one
to force a particular analytic functional form on the data.
The disadvantage however, is that the resulting models are simply solutions to the Jeans 
equation, and so many of them are likely
to be unphysical (corresponding to a non-physical distribution function) or possibly unstable, 
but examining the ($\sim 10^7$)
models individually in the detail necessary to ascertain their
validity would be an immense task, far beyond the scope of this contribution. 

Nevertheless, taking the distribution in Figure~\ref{fig:Jeans_posterior} at face value gives a
99\% upper bound to the dark matter inside of $10\arcmin$ of $1.1\times10^6\Msun$, which is somewhat
higher than the previous method. The strong difference is that the Jeans
equation solution has a 99\% lower limit at $7.5\times 10^5\Msun$, effectively demanding more
dark matter than stars within $10\arcmin$. The difference with respect to
the DM-free models is that that the solutions are better able to follow the central enhancement
of the velocity dispersion. In order to confirm this suspicion, we re-ran the Jeans 
equation simulations from \S\ref{sec:kinematic_models_DM} (that include dark matter),
this time ignoring the most central 25 radial velocity measurements. The result was a solution very similar
to that found in \S\ref{sec:kinematic_models_noDM}, with only a small dark matter
fraction (mean total dark matter mass 
of $1.6\times10^5\Msun$ inside of $10\arcmin$). This experiment shows that the difference in total dark matter
content between the distribution function model solution of \S\ref{sec:dynamical_models_DM} 
and the Jeans equation solution of \S\ref{sec:kinematic_models_DM} is due to the latter 
attempting to fit better the central velocity dispersion, and that it is those central
stars that are driving the requirement for a (small) dark matter fraction.
The MCMC algorithm accommodates this mass by changing primarily the anisotropy profile,
which can be seen to be negative at $r < 0.5\arcmin$ in Figure~\ref{fig:Jeans_composite}d (green dots).

It is interesting to consider which model is preferable: the best
composite Michie plus double power-law model, or the best model
resulting from the Jeans equation analysis? 
The traditional frequentist $\chi^2$ approach is unfortunately not of much
use due to the large number of parameters in the Jeans equation
solver (we would simply find a negative reduced-$\chi^2$, indicating that we have too
many parameters). However, answering this question properly with Bayesian
evidence is very challenging because the parameters used in the two
analysis methods are very different. Indeed, it would require
integrating the models over all parameter values, which would be
extremely computationally costly, and is beyond the scope of the present work.

In this paper we have used kinematics measurements of NGC~2419 to test
the hypothesis that this globular cluster is embedded in a massive DM
halo. Another dark component that can in principle be detected via
kinematics measurements is a central intermediate mass black hole
\citep{Bau05}.  However, the current kinematic sample does not allow one
to set significant constraints on the mass of this putative central
black hole, because it affects only the very central ($\simlt 0\scnd1$)
velocity-dispersion profile, which is barely probed by current
observations.

\subsection{Comparison with previous work}
\label{sec:comp}

\citet{Bau09} assume that the putative halo around NGC~2419 has a NFW profile (with
$\rs=500\pc$) and use their data to limit the dark mass inside of 500\pc\ to
$\Mdm(r<500\pc)<10^7\Msun$. Given their
model parameters, this corresponds to $\Mdm(r<10\arcmin)<3.9\times 10^6 \Msun$.  
In the work of \citet{Con11}, they also assume
that the halo has a NFW profile (with $\rs=250\pc$ and $r_{vir}=1\kpc$)
and conclude that $\Mdm(r<1\kpc)<10^6\Msun$, which, for their model
parameters, gives a very tight $\Mdm(r<10')<2.4\times 10^4\Msun$.  Our results for 
the two component models are in reasonable agreement with 
these previous results. The limits provided above, 
are more stringent than those of \citet{Bau09}, and unlike \citet{Con11}, our
method does not require the cluster to have suffered strong two-body relaxation, which as we have mentioned, 
should not have occurred in NGC~2419.

\begin{table*}
\begin{center}
\caption{Model parameters and derived quantities for the four families of
models investigated in this contribution. Where possible, we report the mean 
value and root mean square dispersion of the MCMC solutions. Due to
the construction of our Jeans equation algorithm, the stellar central density is expensive
to calculate at every MCMC iteration, so we report only the values of
the most likely solution. Furthermore, the posterior probability distribution for the 
dark mass in Fig.~2 is highly skewed, so we only
report the most likely value for that model.}
\label{tab:Michie}
\begin{tabular}{ccccc}
\hline\hline
  & Michie & Jeans without DM & Michie + DM & Jeans with DM \\
\hline
stellar central density ($\Msun \, \pc^{-3}$) & $57.3\pm4.8$ & $74.4$ (best value) & $59.2\pm10$ & $66.9$ (best value)\\
stellar core radius ($\pc$) & $12.6\pm0.6$ & ... & $10.8\pm0.5$ & ... \\
anisotropy radius ($\pc$) & $14.2\pm2.3$ & ... & $18.3\pm2.7$ & ... \\
central potential   & $3.6\pm0.4$ & ... & $4.7\pm1.1$ & ... \\
stellar $M/L$ & $1.72\pm0.11$ & $1.89\pm0.07$ & $1.62\pm0.17$ & $1.15\pm0.12$\\
Stellar mass ($10^5 \Msun$) & $8.1\pm0.2$ & $9.0\pm0.1$ & $7.6\pm0.8$  & $5.5\pm0.6$ \\
Dark mass ($10^5 \Msun$) & $0$ & $0$ & $0.5$ (best value)  & $9.7\pm0.8$ \\
\hline\hline
\end{tabular}
\end{center}
\end{table*}

\section{Conclusions}
\label{sec:con}

We have undertaken an analysis of current data on the kinematics,
structure and stellar mass to light ratio of the
Galactic globular cluster NGC~2419. This distant halo cluster is a
very interesting specimen to study, since it is massive and as such
may have been able to maintain a residual dark matter halo over the
course of its assimilation into the Milky Way. Thus NGC~2419 allows us
to examine the possibility of whether (some) globular clusters formed
along with (their now disrupted?) dwarf galaxy hosts within mini dark
matter halos.

Our earlier analysis (Paper~I) showed that a simple
Michie model gives an excellent fit to the state-of-the-art
measurements of the kinematics and surface brightness profile of the
cluster presented in that contribution. Here we extended that study to investigate whether composite
models with an additional dark matter component in the form of a
generic double power-law could also be accommodated. To this end we
implemented an MCMC parameter search for solutions of this composite
system. The Michie distribution function is found to be
highly restrictive; while it is as yet not possible to completely rule out 
the presence of dark matter in NGC~2419, the preferred dark matter
fraction is 6\% of the stellar mass within 10\arcmin\ (the approximate limit of the cluster),
and there is virtually no room for astrophysically
interesting amounts of dark matter that could hint at the formation
of clusters within dark matter halos.

In a major effort to assess the model-dependence of these results we developed
a new Markov-Chain Monte Carlo algorithm that solves the spherical Jeans equation using 
non-parametric profiles for the density and velocity dispersion profiles.
A very high number of free parameters is required to follow the gradients with sufficient resolution.
The code uses a hierarchy of resolutions so as to probe the parameter space
as efficiently as possible. These ideas (and the software) should be applicable in the future
to analyse other approximately spherical systems of interest, such as dwarf galaxies, elliptical galaxies, or galaxy clusters.

The more general Jeans equation solutions prefer a small dark matter
fraction, approximately twice the stellar mass content, and at face value
rule out the possibility that there is no dark matter. This last inference 
depends, however, on the validity of the conclusion by \citet{Bel12} that the stellar mass to
light ratio is invariant with radius.
For the most likely solution, the inferred dark matter profile has a very different
shape to an NFW model, so it is possible that this putative dark component, if real, 
is related to unexpected stellar remnants (the expected remnants were 
taken into account by \citealt{Bel12}) rather than cosmological dark matter
(note however, that the resolution of present-day simulations that include the relevant baryonic physics
do not allow reliable predictions for the profile of dark matter mini-halos of this mass).

In their analysis, \citet{Bel12} found that assuming a 100\% retention of all the dark remnants 
would have changed the best fit ${\Mstar/L_V}$ from 1.49 to 1.6. Hence in a
standard scenario, changing the fraction of remnant stars will not have a sufficiently
large effect to account for Fig.~\ref{fig:ML}.
On the other hand, as is pointed out in that contribution,
all the currently available (and highly uncertain) models invoked to explain the 
multiple populations in GCs envisage (a) a progenitor of the present-day cluster with a 
baryonic mass a factor $\simgt 10$ larger than today, (b) a preferential loss 
of first generation stars with respect to the second generation, (c) a strong 
difference in the radial distribution of stars of the two generations 
(the second one being more concentrated, as is generally observed; see \citealt{Lar11}).
This complex star formation history and strong variations in the cluster potential must have 
occurred in the first $<500\Myr$ of the cluster life. It can be conceived that at this stage 
both (i) an anomalous production and retention of dark remnants occurred, 
and (ii) a radial distribution of dark remnants different from that of the light 
was produced and then remained un-modified in that non-collisional environment.

Therefore if the observed dark matter is of baryonic origin it is likely that 
it has something to do with the origin and early enrichment of the multiple 
populations \citep{Coh11, Muc12} inhabiting NGC2419. On the other hand one may imagine that 
we see the small remnant of a larger non-baryonic cosmological halo that was 
nearly completely stripped away from the cluster at early times (see, e.g. \citealt{Pen08}). During the first 
approach to its peri-galacticon, it may have provided the potential well that 
allowed the cluster to retain ejecta from first generation stars.

Thus all the models we have explored turn out to have very little dark matter in their central
regions, and we rule out any connection with mini dark matter halos of the mass that
is associated with those in which dwarf galaxies are embedded. This conclusion,
though consistent with earlier work on this particularly promising cluster, and consistent
also with a large literature on more nearby systems (but which on the whole are clearly 
not simple isolated systems), remains puzzling. The connection between globular clusters 
and dwarf galaxy remnants is becoming more and more compelling \citep{Mac10},
so it is natural to expect that some clusters retained some of the dark matter of their former
host. It is possible that this problem will only be resolved conclusively by obtaining a larger
sample of distant halo clusters, but that will require observing further afield in M31,
and to be feasible will require the next generation of instruments.

\section*{Acknowledgements}
R.I. gratefully
acknowledges support from the Agence Nationale de la Recherche though
the grant POMMME (ANR 09-BLAN-0228).
C.N. is supported by the MIUR grant PRIN2008. 
AS acknowledges the support of INAF through the 2010 postdoctoral fellowship grant.
M.B. acknowledges the financial support of INAF through the PRIN-INAF 2009 grant assigned to the project {\em Formation and evolution of massive star clusters}, P.I.: R. Gratton. 

\section*{Apendix A}

The distribution function proposed by \citet{Mic63} has the following phase-space form:
\begin{eqnarray}
\label{eq_df}
f(E,L)&=&\fzero~\exp\left(-\frac{L}{2\sigmaK^{2}\ra^{2}}\right)\left[\exp\left(-\frac{E}{\sigmaK^{2}}\right)-1\right],\nonumber\\
f(r,\vr,\vt)&=&\fzero~\exp\left[-\frac{\vt^{2}}{2\sigmaK^{2}}\left(\frac{r}{\ra}\right)^{2}\right]~\times\nonumber\\
& &\left[\exp\left(-\frac{\vr^{2}+\vt^{2}}{2\sigmaK^{2}}-\frac{\psi}{\sigmaK^{2}}\right)-1\right],
\end{eqnarray}
where $E$ and $L$ are the energy and angular momentum per unit mass,
respectively, $\psi$ is the effective potential defined as the
difference between the cluster potential (including the contribution
of both stars and the dark matter --- the latter component will be added 
in Section~\ref{sec:dynamical_models_DM}) 
at a given radius $r$ and the potential at the
cluster tidal radius ($\psi=\phi-\phit$), $\fzero$ is a
normalization coefficient factor, $\sigmaK$ is a reference velocity
dispersion, $\vt$ and $\vr$ are the radial and tangential component of
the velocity and $\ra$ is the anisotropy radius: the velocity
distribution tends to be isotropic at $r<\ra$ and radially anisotropic
at $r>\ra$.

The density and the radial and tangential components of the velocity
dispersion can be derived by integrating the above distribution
function:
\begin{eqnarray}
\rhostar(r)&=& 4 \pi \int_{0}^{\sqrt{-2\psi}}\int_{0}^{\sqrt{-2\psi-\vr^{2}}}
\vt~f(r,\vr,\vt) ~\mathrm{d} \vt \mathrm{d} \vr,\nonumber\\
\sigmar^{2}(r)&=&\frac{4
\pi}{\rhostar(r)}\int_{0}^{\sqrt{-2\psi}} \vr^{2} \int_{0}^{\sqrt{-2\psi-\vr^{2}}} \vt~ f(r,\vr,\vt)
~\mathrm{d} \vt \mathrm{d} \vr,\nonumber\\
\sigmat^{2}(r)&=&\frac{4 \pi}{\rhostar(r)}\int_{0}^{\sqrt{-2\psi}}
\int_{0}^{\sqrt{-2\psi-\vr^{2}}} \vt^{3}~f(r,\vr,\vt)
~\mathrm{d} \vt \mathrm{d} \vr.\nonumber
\end{eqnarray}
It is convenient to work with the dimensionless quantities
\begin{eqnarray}
\zeta=\frac{\vr^2}{2\sigmaK^{2}}, & \eta=\frac{\vt^2}{2\sigmaK^{2}},\nonumber\\ 
\rhostartilde=\frac{\rhostar}{\rhostarzero}, & \tilde{r}=\frac{r}{\rc},\nonumber\\
W=-\frac{\psi}{\sigmaK^{2}}, & \ratilde=\frac{\ra}{\rc},\nonumber
\end{eqnarray}
where $\rhostarzero$ is the central cluster density and
$$\rc\equiv\left(\frac{9 \sigmaK^{2}}{4 \pi G \rhostarzero}\right)^{1/2}$$
is the core radius (King 1966). The above equations can be then written 
as
\begin{eqnarray}
\label{eq_models}
\rhostartilde(\tilde{r})&=&\frac{\int_{0}^{W}\zeta^{-\frac{1}{2}}\int_{0}^{W-\zeta}e^{-\frac{\eta\tilde{r}^{2}}{\ratilde^{2}}}(e^{W-\eta-\zeta}-1) 
~\mathrm{d}\eta ~\mathrm{d}\zeta}{\int_{0}^{\Wzero}\zeta^{-\frac{1}{2}}\int_{0}^{\Wzero-\zeta}e^{-\frac{\eta\tilde{r}^{2}}{\ratilde^{2}}}(e^{\Wzero-\eta-\zeta}-1) 
~\mathrm{d}\eta ~\mathrm{d}\zeta},\nonumber\\
\sigmar^{2}(\tilde{r})&=&\frac{2\sigmaK^{2}\int_{0}^{W}\zeta^{\frac{1}{2}}\int_{0}^{W-\zeta}
e^{-\frac{\eta\tilde{r}^{2}}{\ratilde^{2}}}(e^{W-\eta-\zeta}-1) 
~\mathrm{d}\eta ~\mathrm{d}\zeta}{\int_{0}^{W}\zeta^{-\frac{1}{2}}\int_{0}^{W-\zeta}e^{-\frac{\eta\tilde{r}^{2}}{\ratilde^{2}}}(e^{W-\eta-\zeta}-1) 
~\mathrm{d}\eta ~\mathrm{d}\zeta},\nonumber\\
\sigmat^{2}(\tilde{r})&=&\frac{2\sigmaK^{2}\int_{0}^{W}\zeta^{-\frac{1}{2}}\int_{0}^{W-\zeta}
\eta e^{-\frac{\eta\tilde{r}^{2}}{\ratilde^{2}}}(e^{W-\eta-\zeta}-1) 
~\mathrm{d}\eta ~\mathrm{d}\zeta}{\int_{0}^{W}\zeta^{-\frac{1}{2}}\int_{0}^{W-\zeta}e^{-\frac{\eta\tilde{r}^{2}}{\ratilde^{2}}}(e^{W-\eta-\zeta}-1) 
~\mathrm{d}\eta ~\mathrm{d}\zeta},\nonumber\\
\end{eqnarray}
which can be solved once the potential $W(\tilde{r})$ is known.  This
last quantity can be written as the sum of the stellar and DM
contributions $W=\Wstar+\Wdm$. Both contributions must obey
independently the Poisson equation
\begin{eqnarray}
\label{eq_pois}
\grad^{2}\phi_{i}&=&4 \pi G \rho_{i},\nonumber\\
\frac{\d^{2}W_{i}}{\d\tilde{r}^{2}}+\frac{2}{\tilde{r}}\frac{\d W_{i}}{d\tilde{r}}&=&-9\tilde{\rho_{i}} \, . \nonumber\\
\end{eqnarray}
Here, the subscript $i$ refers to either stars or DM. 
For the stellar component we leave
$\Wzero$ as a free parameter and solve equation~(\ref{eq_pois})
adopting the boundary conditions at the center
\begin{eqnarray}
\label{bound_eq}
\Wstarzero&=& \Wzero-\Wdmzero \, , \nonumber\\
\frac{\d \Wstarzero}{\d\tilde{r}}&=&0 \, .\nonumber
\end{eqnarray}

Equations~(\ref{eq_models}) and (\ref{eq_pois}) can be integrated
numerically to derive the three-dimensional density profile, and the
radial and tangential velocity dispersion profiles.  Finally, the
above profiles have been projected on the plane of the sky to obtain
the surface mass density
\begin{equation}
\Sigmastar(R)=2\int_{R}^{\rttilde}
\frac{\rhostartilde\tilde{r} \mathrm{d} \tilde{r}}{\sqrt{\tilde{r}^{2}-R^{2}}}\nonumber
\end{equation}
and the line-of-sight velocity dispersion
\begin{equation}
\sigmaV^{2}(R)=\frac{1}{\Sigmastar(R)}\int_{R}^{\rttilde}
\frac{\rhostartilde\left[2\sigmar^{2}\left(\tilde{r}^{2}-R^{2}\right)+\sigmat^{2}R^{2}\right] \mathrm{d} \tilde{r}}{\tilde{r}\sqrt{\tilde{r}^{2}-R^{2}}} \, .\nonumber
\end{equation}

As described in Section~\ref{sec:dynamical_models_noDM}, the full
likelihood analysis performed here requires for each model the
knowledge of the distribution of projected velocities (not only its
second-order moment).  This function can be calculated as
\begin{eqnarray}
&&f_p(v,R)=\int_{0}^{\rt}\frac{\rhostar(r) r}{\sqrt{r^{2}-R^{2}}}\int_{0}^{\sqrt{-2\psi-v^2}}\nonumber\\
&&\times\int_{0}^{\sqrt{-2\psi-v^2-\vx^2}} f(r,\vr',\vt')~\d \vy \d \vx \d r \, ,
\end{eqnarray}
adopting the following change of variables
\begin{eqnarray}
\vr'&=&v~\sin\alpha +\vx~\cos\alpha,\nonumber\\
\vt'&=&[(v~\cos\alpha -\vx~\sin\alpha)^{2}+\vy^{2}]^{1/2} \, ,\nonumber
\end{eqnarray}
where $\alpha=\arccos(R/r)$.


\begin{thebibliography}{}
\bibitem[\protect\citeauthoryear{Amorisco 
\& Evans}{2011}]{Amo11}Amorisco N.~C., Evans N.~W., 2011, MNRAS, 411, 2118
\bibitem[\protect\citeauthoryear{Baumgardt, Makino, \& Hut}{2005}]{Bau05}
Baumgardt H., Makino J., Hut P., 2005, ApJ, 620, 238
\bibitem[\protect\citeauthoryear{Baumgardt \& Mieske}{2008}]{Bau08}
Baumgardt H., Mieske S., 2008, MNRAS, 391, 942 
\bibitem[\protect\citeauthoryear{Baumgardt et al.}{2009}]{Bau09}
Baumgardt H., C{\^o}t{\'e} P., Hilker M., Rejkuba M., Mieske S., Djorgovski S.~G., Stetson P., 2009, MNRAS, 396, 2051 
\bibitem[\protect\citeauthoryear{Bekki \& Yong}{2012}]{Bek12} 
Bekki, K., Yong, D., 2012, MNRAS, 419, 2063
\bibitem[\protect\citeauthoryear{Bellazzini et al.}{2012}]{Bel12} 
Bellazzini, M., Dalessandro, E., Sollima, A., Ibata, R., 2012, MNRAS, 423, 844
\bibitem[\protect\citeauthoryear{Bellazzini et al.}{2008}]{Bel08} 
Bellazzini M., Ibata, R., Chapman, S., Mackey, D., Monaco, L., Irwin, M., Martin, N., Lewis, G., Dalessandro, E., 2008, AJ, 136, 1147
\bibitem[Binney \& Mamon(1982)]{Bin82}
Binney J., Mamon G.A., 1982, MNRAS, 200, 361
\bibitem[Binney \& Tremaine(2008)]{Bin08}
Binney J., Tremaine S., 2008, Galactic Dynamics 2nd Ed., Princeton University Press, Princeton 
\bibitem[\protect\citeauthoryear{Bradford et al.}{2011}]{Bra11}
Bradford J.~D., et al., 2011, ApJ, 743, 167
\bibitem[\protect\citeauthoryear{Burkert}{1995}]{Bur95}
Burkert A., 1995, ApJ, 447, L25 
\bibitem[\protect\citeauthoryear{Ciotti \& Morganti}{2010a}]{Cio10a}
Ciotti L., Morganti L., 2010a, MNRAS, 401, 1091
\bibitem[\protect\citeauthoryear{Ciotti \& Morganti}{2010b}]{Cio10b}
Ciotti L., Morganti L., 2010b, MNRAS, 408, 1070
\bibitem[\protect\citeauthoryear{Cohen, Huang \& Kirby}{2011}]{Coh11}
Cohen, J., Huang, W., Kirby, E., 2011, ApJ, 740, 60
\bibitem[\protect\citeauthoryear{Conroy, Loeb, \& Spergel}{2011}]{Con11}
Conroy C., Loeb A., Spergel D., 2011, ApJ, 741, 72
\bibitem[\protect\citeauthoryear{Dalessandro et al.}{2008}]{Dal08}
Dalessandro, E., Lanzoni, B., Ferraro, F., Vespe, F., Bellazzini, M. Rood, R., 2008, ApJ, 681, 311
\bibitem[\protect\citeauthoryear{Di Criscienzo et al.}{2011}]{DiC11}
Di Criscienzo M., et al., 2011, AJ, 141, 81
\bibitem[\protect\citeauthoryear{Fridman 
\& Polyachenko}{1984}]{Fri84}Fridman A.~M., Polyachenko V.~L., 1984, Physics of Gravitating Systems. Springer, New York
\bibitem[\protect\citeauthoryear{Goodman \& Weare}{2010}]{Goo10}
Goodman, J., Weare, J., 2010, Comm. App. Math. Comp. Sci., 5, 65
\bibitem[\protect\citeauthoryear{Gregory}{2005}]{Gre05}
Gregory, P., 2005, Bayesian Logical Data Analysis for the Physical Sciences. Cambridge University Press
\bibitem[\protect\citeauthoryear{Harris}{1996}]{Har10}
Harris W.~E., 1996, AJ, 112, 1487
\bibitem[\protect\citeauthoryear{Heggie \& Hut}{1996}]{Heg96}
Heggie D.~C., Hut P., 1996, IAUS, 174, 303 
\bibitem[\protect\citeauthoryear{Ibata et al.}{2011a}]{Iba11a} 
Ibata R., Sollima A., Nipoti C., Bellazzini M., Chapman S.~C., Dalessandro E., 2011a, ApJ, 738, 186
\bibitem[\protect\citeauthoryear{Ibata et al.}{2011b}]{Iba11b} 
Ibata R., Sollima A., Nipoti C., Bellazzini M., Chapman S., Dalessandro E., 2011b, ApJ, 743, 43
\bibitem[\protect\citeauthoryear{Lardo et al.}{2011}]{Lar11} 
Lardo, C., Bellazzini M., Pancino, E., Carretta, E., Bragaglia, A., Dalessandro E., 2011, A\&A, 525, 114
\bibitem[\protect\citeauthoryear{Mackey et al.}{2010}]{Mac10} 
Mackey, D., Huxor, A., Ferguson, A., Irwin, M., Tanvir, N., McConnachie, A., Ibata, R., Chapman, S., Lewis, G., 2010, ApJ, 717, 11
\bibitem[\protect\citeauthoryear{Mapelli et al.}{2009}]{Map09}
Mapelli M., Ripamonti E., Battaglia G., Tolstoy E., Irwin M.~J., Moore B., Sigurdsson S., MNRAS, 396, 1771
\bibitem[\protect\citeauthoryear{Mashchenko \& Sills}{2005a}]{Mas05a}
Mashchenko S., Sills A., 2005a, ApJ, 619, 243
\bibitem[\protect\citeauthoryear{Mashchenko \& Sills}{2005b}]{Mas05b}
Mashchenko S., Sills A., 2005b, ApJ, 619, 258
\bibitem[\protect\citeauthoryear{Merritt}{1985}]{Mer85} 
Merritt D., 1985, AJ, 90, 1027 
\bibitem[\protect\citeauthoryear{Michie}{1963}]{Mic63}
Michie R.~W., 1963, MNRAS, 125, 127
\bibitem[\protect\citeauthoryear{Milgrom}{1983}]{Mil83} 
Milgrom M., 1983, ApJ, 270, 365 
\bibitem[\protect\citeauthoryear{Milgrom}{1986}]{Mil86} 
Milgrom M., 1986, ApJ, 306, 9 
\bibitem[\protect\citeauthoryear{Moore}{1996}]{Moo96}
Moore B., 1996, ApJ, 461, L13 
\bibitem[\protect\citeauthoryear{Mucciarelli et al.}{2012}]{Muc12}
Mucciarelli, A., Bellazzini, M., Ibata, R., Merle, T., Chapman, S., Dalessandro, E., Sollima, A., 2012, arXiv:1208.0195
\bibitem[\protect\citeauthoryear{Navarro \& Steinmetz}{2000}]{Nav00}
Navarro J.~F., Steinmetz, M., 2000, ApJ, 538, 477
\bibitem[\protect\citeauthoryear{Navarro, Frenk, \& White}{1996}]{Nav96}
Navarro J.~F., Frenk C.~S., White S.~D.~M., 1996, ApJ, 462, 563 (NFW)
\bibitem[\protect\citeauthoryear{Nipoti, Londrillo, \& Ciotti}{2002}]{Nip02}
Nipoti C., Londrillo P., Ciotti L., 2002, MNRAS, 332, 901 
\bibitem[\protect\citeauthoryear{Nipoti, Ciotti, \& Londrillo}{2011}]{Nip11}
Nipoti C., Ciotti L., Londrillo P., 2011, MNRAS, 414, 3298 
\bibitem[\protect\citeauthoryear{Osipkov}{1979}]{Osi79}
Osipkov L.~P., 1979, Soviet Astron. Lett., 5, 42
\bibitem[\protect\citeauthoryear{Peebles}{1984}]{Pee84}
Peebles P.~J.~E., 1984, ApJ, 277, 470 
\bibitem[\protect\citeauthoryear{Pe\~{n}arrubia, Navarro \& McConnachie}{2008}]{Pen08}
Pe\~{n}arrubia, J., Navarro, J., McConnachie, A., 2008, ApJ, 673, 226
\bibitem[\protect\citeauthoryear{Polyachenko 
\& Shukhman}{1981}]{Pol81}Polyachenko V.~L., Shukhman I.~G., 1981, SvA, 25, 533 
\bibitem[\protect\citeauthoryear{Quinlan, Hernquist \& Sigurdsson}{1995}]{Qui95} 
Quinlan G.~D., Hernquist L., Sigurdsson S., 1995, ApJ, 440, 554
\bibitem[\protect\citeauthoryear{Saitoh et al.}{2006}]{Sai06} 
Saitoh T.~R., Koda J., Okamoto T., Wada K., Habe A., 2006, ApJ, 640, 22
\bibitem[\protect\citeauthoryear{Sanders}{2012}]{San12} 
Sanders, R., 2012, MNRAS, 419, 6
\bibitem[\protect\citeauthoryear{Sollima, Bellazzini \& Lee}{2012}]{Sol12} 
Sollima, A., Bellazzini, M., Lee, J.-W., 2012, arXiv:1206.4828
\bibitem[\protect\citeauthoryear{Walker}{2012}]{Wal12} 
Walker, 2012, arXiv:1205.0311
\bibitem[\protect\citeauthoryear{Zocchi, Bertin, \& Varri}{2012}]{Zoc12} 
Zocchi A., Bertin G., Varri A.~L., 2012, A\&A, 539, A65
\end{thebibliography}
\end{document}